\documentclass[a4paper,fleqn]{cas-sc}

\usepackage[numbers]{natbib}

\usepackage{amsthm}
\usepackage{float}
\usepackage[normalem]{ulem}
\usepackage{soul,color}
\usepackage{tablefootnote,gensymb}
\usepackage{eurosym}
\usepackage{tabularx}
\usepackage{svg}
\usepackage{enumitem}
\usepackage{algorithm}
\usepackage{algpseudocode}
\usepackage{textcomp}
\usepackage{tikz}
\usepackage{microtype}

\usepackage{caption}

\usepackage{subcaption}
\DeclareSymbolFont{Xlargesymbols}{OMX}{cmex}{m}{n}
\DeclareMathSymbol{\Xsum}{\mathop}{Xlargesymbols}{80}

\def\BibTeX{{\rm B\kern-.05em{\sc i\kern-.025em b}\kern-.08em
    T\kern-.1667em\lower.7ex\hbox{E}\kern-.125emX}}

\usepackage{tikz}
\usepackage{eso-pic}

\newcommand{\PL}[1]{\textcolor{black}{#1}}
\newcommand{\Ahmad}[1]{\textcolor{black}{#1}}

\begin{document}
\let\WriteBookmarks\relax
\def\floatpagepagefraction{1}
\def\textpagefraction{.001}
\shorttitle{Anomaly Detection in Offshore Open RAN Using LSTM Models on a Novel AI-Driven Cloud-Native Data Platform}
\shortauthors{Ahmad \& Li et~al.}

\title [mode = title]{Anomaly Detection in Offshore Open Radio Access Network Using Long Short-Term Memory Models on a Novel Artificial Intelligence-Driven Cloud-Native Data Platform}                      
\author[1]{Abdelrahim Ahmad}[orcid=https://orcid.org/0000-0002-6980-5267]
\cormark[1]
\ead{Abdelrahim.Ahmad@boldyn.com}


\address[1]{Data Science and Data Engineering Section, Department of Technology, Boldyn Networks, London W2 6BD, United Kingdom}
\author[2]{Peizheng Li}[orcid=https://orcid.org/0000-0003-1516-1993]

\ead{lipeizheng16@gmail.com}

\address[2]{Department of Electrical and Electronic Engineering, University of Bristol, Bristol BS8 1UB, United Kingdom}
\author[2]{Robert Piechocki}[orcid=https://orcid.org/0000-0002-4879-1206]
\ead{R.J.Piechocki@bristol.ac.uk}


\author[1]{Rui Inacio}[]
\ead{Rui.Inacio@boldyn.com}

\cortext[cor1]{Corresponding author}

\begin{abstract}
The Radio Access Network (RAN) is a critical component of modern telecommunications infrastructure, currently evolving towards disaggregated and open architectures. These advancements are pivotal for integrating intelligent, data-driven applications aimed at enhancing network reliability and operational autonomy through the introduction of cognitive capabilities, as exemplified by the emerging Open Radio Access Network (O-RAN) standards.
Despite its potential, the nascent nature of O-RAN technology presents challenges, primarily due to the absence of mature operational standards. This complicates the management of data and intelligent applications, particularly when integrating with traditional network management and operational support systems. Divergent vendor-specific design approaches further hinder migration and limit solution reusability. These challenges are compounded by a skills gap in telecommunications business-oriented engineering, which remains a key barrier to effective O-RAN deployment and intelligent application development.
\Ahmad{To address these challenges, Boldyn Networks developed a novel cloud-native data analytics platform, specifically designed to support scalable AI integration within O-RAN deployments. This platform underwent rigorous testing in real-world scenarios, and applied advanced Artificial Intelligence (AI) techniques to improve operational efficiency and customer experience. Implementation involved adopting Development Operations (DevOps) practices, leveraging data lakehouse architectures tailored for AI applications, and employing sophisticated data engineering strategies.}
\PL{The platform successfully addresses connectivity challenges inherent in real-world offshore windfarm deployments using Long Short-Term Memory (LSTM) models for anomaly detection in network connectivity. After integrating the LSTM models into the network control, more than 90 percent of connectivity issues were reduced in runtime. This marks a step toward autonomous, self-organising, and self-healing networks.}
\end{abstract}


\AddToShipoutPictureFG*{%
  \begin{tikzpicture}[remember picture, overlay]
    \node[anchor=north, minimum width=\paperwidth,
          fill=white,text=black,font=\normalsize,align=center]
      at (current page.north) {This work has been published in \textit{Engineering Applications of Artificial Intelligence}, \\
Volume 161, Part C, 12 December 2025, 112274. \\ 
https://doi.org/10.1016/j.engappai.2025.112274};
  \end{tikzpicture}%
}

\maketitle
\section{Introduction}
\label{sec:introduction}
Telecommunication networks are essential to many aspects of our lives, driving digital transformation and revolutionizing communication. The benefits of these networks are numerous. Recently, their importance has surged due to the proliferation of various types of user equipment (UE), internet of things (IoT) devices, autonomous operations, and services that require faster, more reliable, resilient, secure, and private connectivity. This has led to a substantial rise in the demand for private networks, amplifying the challenges of managing many smaller, tailored mobile networks to deliver high-quality services.

To address these escalating demands, innovative enhancements in network design have emerged. A recent progress in this arena is the advent of open radio access network (O-RAN) technology. O-RAN aims to disaggregate the monolithic, single-vendor RAN reducing infrastructure costs and vendor lock-in, and paving the way for network programmability, ultimately leading to autonomous network operations by leveraging data and native-supported in-network artificial intelligence (AI) techniques, aiming to streamline the complexities of designing, delivering, managing and operating private networks. It establishes a new framework of standards and principles for wireless networking, emphasizing open standards, interfaces, functions, and interoperability to foster greater market competition~\cite{OpenRANyang}.

A wide range of applications are being investigated in O-RAN due to their critical role in improving system functionality, such as predictive maintenance and anomaly detection, energy efficiency optimization, automated network configuration and healing, improved quality of service (QoS) and traffic management, enhanced user admission control, dynamic RAN slicing~\cite{li2023digital,frcmn.2023.1127039,polese2023understanding,10329948}. AI and machine learning (ML) approaches are considered the main tools that can address these challenges~\cite{singh2020evolution}.

However, O-RAN is still a relatively new approach and developing solutions in it comes with many realistic implementation challenges, such as interoperability with legacy network management systems, big data problem, data integration issues, the need for sophisticated pipelines for data acquisition and processing to produce AI- and data-driven applications, as well as management and other technical complexities. In addition to these challenges, there is a shortage of experienced engineers and a growing need for engineering roles with specific skill sets that are essential to support the transformation of the O-RAN architecture.

\Ahmad{A clear example of these challenges was observed during the deployment of a live offshore private network supporting windfarm vessels in the North Sea. The network, based on O-RAN principles, operated in a complex environment shaped by changing weather conditions, vessel mobility, and the wide area of coverage in the sea. These factors made connectivity management more difficult and resulted in recurring connectivity issues. Existing tools offered limited visibility and control, making it difficult to identify and address the root causes. Addressing this required more than just an AI model, it required a complete rethinking of how data is collected, processed, and used across the network to support intelligent and automated decision-making. Building a dedicated platform to host data and develop intelligent applications was a critical first step toward solving this complex challenge. It provided the foundation for a sustainable and scalable approach to deploying AI in O-RAN environments. Such platforms are not readily available in the market and remain in the early stages of maturity, particularly for multi-vendor, production-grade telecom networks. Therefore, developing this platform was not just a support element, it was a necessary innovation that made it possible to manage complexity, enable automation, and ensure long-term adaptability of AI solutions in O-RAN.} Conversely, relying on traditional data solutions from a single vendor is nearly impossible in a multi-vendor environment. These solutions are often limited in scope, requiring significant customization and additional tools to meet operational needs, ultimately leading to higher costs and complexity.

\begin{figure*}
\centering
    \includegraphics[width=1\textwidth]{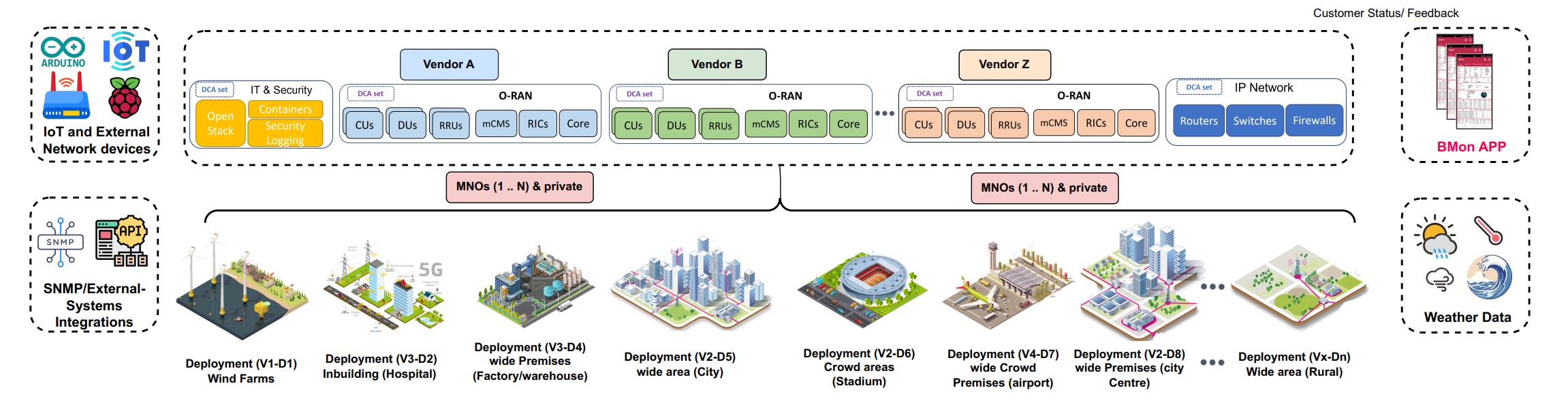}
    \caption{Multi-vendor O-RAN network setup with Vendor-A in light blue, Vendor-B in green, and Vendor-Z in orange. In addition to other systems, equipment, and user devices that are working within the network and produce important data.}
    \label{fig:ORAN_architecture}
\end{figure*}

In this paper, we introduce a novel cloud-native, open data analytics platform for O-RAN to address these challenges and streamline the integration of AI applications into the O-RAN management stack. This platform leverages cutting-edge engineering technologies and concepts such as data lakehouse~\cite{armbrust2021lakehouse}, DevOps~\cite{ebert2016devops}, and MLOps. The platform is designed to support the growing number of private network deployments, regardless of the vendor, and the increasing complexity of their use cases. It is built to scale and handle the large amounts of data these networks generate.
The data sources collected in the platform are big, variant and complex, coming from many different systems, as shown in Fig.~\ref{fig:ORAN_architecture}. \PL{ Leveraging this data, we applied a long short-term memory (LSTM)-based anomaly detection model to address the critical offshore connectivity challenge. 
By deploying the model in real time, the system enabled proactive anomaly detection and triggered automated mitigation actions, resulting in over 90\% reduction in modem disconnections compared to the conventional method. This deployment represents one of the first real-world, production-grade demonstrations of AI-driven anomaly detection in a live, multi-vendor O-RAN network.}

The main contributions of this work are summarized below and explained in more detail in the literature review (Sec.~\ref{sec:literature_review}), where the novelty of our approach is also highlighted in relation to existing research.

\begin{enumerate}
    \item To the best of the authors’ knowledge, we propose the first holistic, cloud-native data platform architecture for multi-vendor O-RAN environments, tightly aligned with O-RAN principles and designed to support scalable AI lifecycle integration. The platform also addresses gaps in standardization and engineering workflows for deploying AI in telecom networks.
    \item The platform fully automates infrastructure setup, data handling, and AI pipelines using DevOps and GitOps technologies, significantly reducing operational workload and improving development efficiency.
    \item We demonstrate the resolution of a real-time offshore connectivity problem in production using an LSTM-based anomaly detection model deployed within the platform, showcasing both technical feasibility and operational impact.
\end{enumerate}

\Ahmad{The rest of this paper is organized as follows. Sec.~\ref{sec:literature_review} introduces the background of O-RAN and how AI is being used to make networks more intelligent, along with a review of related work and the unique contributions and novelty of this study. Sec.~\ref{sec:problem_statement} presents the real-world problem faced in offshore deployments and explains why AI was chosen to solve it and why the data platform was crucial to do that. Sect.~\ref{sec:data_platform} describes the design of the proposed cloud-native data platform, focusing on its architecture, key technologies, component layers, and how it supports standardizing the AI development. Sect.~\ref{sec:proposed_solution} details the AI-based anomaly detection use case, including model design, data preparation and feature engineering, deployment, and performance results. Sec.~\ref{sec:discussions} discusses current limitations and future improvements, and Sec.~\ref{sec:conclusion} concludes the paper by summarizing the main findings and impact of this work.}

\section{Literature review: AI for O-RAN}
\label{sec:literature_review}
\PL{In this section, we present the background information about the O-RAN technique, the relevant progress in research, and AI applications of O-RAN.}
\subsection{O-RAN background}

The RAN is a critical component of a typical mobile communication network, enabling UE to connect to the core network, which then delivers services to users. The evolution of wireless communication systems from First-Generation (1G) to Fifth-Generation (5G) highlights increasing modularity and virtualization of network functionalities. Key advancements in RAN architecture include distributed RAN (D-RAN), centralized (or Cloud) RAN (C-RAN), and virtual RAN (vRAN). The distinctions among these architectures are detailed in~\cite{faisal_2021}.

In the 3GPP 5G new radio (NR) specifications, the traditional base station (BS) is composed of three main components: the centralized unit (CU), distributed unit (DU), and radio unit (RU). The CU and DU together perform the functions of the baseband unit (BBU), while the RU is responsible for signal conversion and radio frequency (RF) transmission.

O-RAN aims to address vendor lock-in issues by promoting the decoupling of hardware and software. This approach advocates for open, standardized interfaces, virtualized network elements, and white-box hardware, driven by principles of intelligence and openness. By doing so, O-RAN seeks to transform the RAN industry, fostering a more flexible and interoperable ecosystem.

\begin{figure}[!h]
    \centering
    \includegraphics[width=0.45\textwidth]{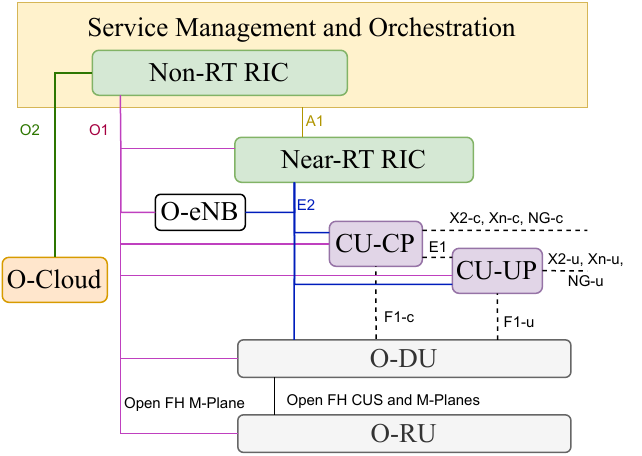}
    \caption{The detailed O-RAN architecture and components.}
    \label{fig:oran architecture}
\end{figure}

Openness in O-RAN involves adopting standardized interfaces to ensure interoperability, enabling seamless integration of hardware components from various vendors, and fostering a multi-vendor RAN ecosystem. The O-RAN Alliance has issued various specifications to support this initiative.
Technically, O-RAN adheres to 3GPP 5G NR specifications, featuring the CU, DU, and RU. As illustrated in Fig.~\ref{fig:oran architecture}, the RU and DU are disaggregated based on the 7.2x split~\cite{O_RAN2023} and connected via the open fronthaul interface. Further segmentation of the CU results in two logical components: the CU control plane (CU-CP) and the CU user plane (CU-UP), enhancing deployment flexibility and reducing latency concerns.
The DU and CU are interconnected through the open midhaul F1 interface, which is divided into F1-C for control plane communications and F1-U for user plane connectivity. 

The intelligence of O-RAN is a pivotal aspect that enhances its functionality through the integration of AI and ML. These advanced technologies enable sophisticated network automation, allowing for dynamic resource allocation, efficient management, and proactive orchestration of network functions and resources. At the heart of this intelligence is RICs~\cite{O_RAN2024}, which are designed to host various applications that drive network optimization and network operational and maintenance processes. RICs are categorized into non-real-time (non-RT) RIC and near-real-time (near-RT) RIC, each supporting different types of applications known as rApps and xApps, respectively. It can be seen from Fig.~\ref{fig:oran architecture} that the near-RT RIC connects to the O-CU/O-DU via the E2 interface for near-real-time control, while the non-RT RIC communicates with the near-RT RIC through the A1 interface for non-real-time control and AI/ML model updates. Additionally, the O1 interface links the non-RT RIC with other RAN components for overall service management and orchestration~\cite{10453148}.
This layered approach ensures that O-RAN can adapt to varying network demands and conditions in real-time, significantly improving performance, reducing operational costs, and enhancing user experience.

\subsection{Literature review}
\Ahmad{Notably, recent advances in AI have led to increased exploration of its applications in communication networks, particularly through O-RAN architectures. Despite growing research interest, most studies are still based on simulations or lab setups. Deploying AI in real, production-grade O-RAN networks remains difficult because the technology is still new, and there are many technical and operational challenges to overcome.}

\Ahmad{For instance, Bonati \textit{et al.}~\cite{BONATI2023109502} introduced the OpenRAN Gym platform, which facilitates the development and testing of AI/ML algorithms on programmable RAN interfaces; however, their work is limited to testbed emulations rather than operational networks. Polese \textit{et al.}~\cite{Polese10024837} offered a comprehensive architectural and algorithmic survey of O-RAN’s intelligent functions, including the roles of RICs and xApps, but did not include any empirical system deployment. Similarly, Yeh \textit{et al.}~\cite{Yeh10335921} and Nagib \textit{et al.}~\cite{Nagib10329948} explored deep learning approaches for network slicing and traffic optimization within O-RAN settings, yet their implementations relied on idealized, offline datasets. Erdol \textit{et al.}~\cite{Hakan10012789} proposed federated learning for traffic steering across distributed O-RANs, emphasizing privacy and coordination, but their framework has not been validated in real deployment scenarios. In~\cite{11005418}, a deep reinforcement learning and graph neural network combined method is introduced to solve the baseband placement problem of O-RAN under different scenarios, but still it is based on a simulation environment.}

\Ahmad{Basaran and Dressler~\cite{BASARAN2025111145} addressed the explainability challenge through their XAInomaly framework, employing deep contractive autoencoders to detect anomalies in synthetic traffic data. While their focus on interpretability is valuable, the system lacks real-time capability and remains confined to simulated environments. Xavier \textit{et al.}~\cite{Xavier10279349} tackled network security through supervised learning models for cyberattack detection in lab-based O-RAN testbeds. Despite achieving high accuracy, their study focuses solely on threat classification and not continuous anomaly detection in operational systems.}

\Ahmad{To date, no published studies have demonstrated the deployment of AI-driven models in live O-RAN environments using real network data, especially in business-critical settings like offshore networks, where connectivity issues can directly impact operations and safety. The present work addresses this critical gap by showcasing the design, implementation, and production deployment of an AI-enabled anomaly detection system within a live offshore mobile network built on O-RAN principles.}
\PL{Table~\ref{table:comparative summary} represents the most relevant recent research efforts and compares them with our contribution. What distinguishes this study is its integration of multiple advanced technologies such AI, big data processing, and cloud-native engineering within a multi-vendor, operational telecom infrastructure. A unified platform was developed in-house to support large-scale telemetry ingestion, real-time analytics, and continuous integration and deployment (CI/CD) workflows tailored specifically for the telemetry and control interfaces of O-RAN. Moreover, this solution uniquely incorporates live network data and UE feedback captured through vessel-based customer-premise equipment (CPE), offering a distinctive, UE perspective view of network performance and enabling faster, more accurate anomaly detection. This convergence of real-world deployment, cross-layer data fusion, and autonomous response positions the work as a novel and timely contribution to the field of intelligent and resilient mobile networks.}

\begin{table}[htbp]
\centering
\footnotesize
\caption{\Ahmad{Comparative Summary of Related Works}}
\renewcommand{\arraystretch}{1.2}
\begin{tabular}{|>{\raggedright\arraybackslash}p{1.8cm}%
                |>{\raggedright\arraybackslash}p{1.6cm}%
                |>{\raggedright\arraybackslash}p{2.0cm}%
                |>{\centering\arraybackslash}p{1.5cm}%
                |>{\raggedright\arraybackslash}p{1.7cm}%
                |>{\centering\arraybackslash}p{1.0cm}%
                |>{\centering\arraybackslash}p{1.0cm}%
                |>{\raggedright\arraybackslash}p{2.2cm}|}
\hline
\textbf{Study \& Reference} & \textbf{Deployment Environment} & \textbf{Data Source} & \textbf{Proposed AI-Data Platform } & \textbf{AI Model} & \textbf{Data Type} & \textbf{Real-Time} & \textbf{Key Outcomes} \\
\hline
Ahmad et al., 2025 \textit{(This work)} & Production Offshore windfarms Network & Real-world FCAPS \& CPE telemetry & Yes, (Cloud-native, multi-vendor) & LSTM & Network + CPE & Yes & 90\% reduction in disconnections \\
\hline
Bonati et al., 2023 \textit{(OpenRAN Gym)} & Simulated testbed & Synthetic data & No & Reinforcement Learning & Network only & Limited & Demonstrated RIC optimization \\
\hline
Basaran \& Dressler, 2025 \textit{(XAI anomaly)} & Simulated O-RAN & Synthetic traffic data & No & Deep Contractive Autoencoder & Network only & No & Introduced explainable anomaly detection \\
\hline
Xavier et al., 2023 \textit{(Early Detection)} & Lab testbed & Simulated attack scenarios & No & ML (various) & Network only & Near real-time & Achieved 95\% accuracy in attack detection \\
\hline
\end{tabular}
\label{table:comparative summary}
\end{table}

\subsection{Summary of the research novelty}

\Ahmad{This research presents a complete and practical innovation in the integration of AI and cloud-native architectures within real-world O-RAN deployments. The work begins with a rare and operationally critical business problem: maintaining stable connectivity in a private O-RAN network deployed across mobile offshore windfarm vessels. These networks face severe challenges, including unstable radio conditions, dynamic weather, signal interference over water, and vessel mobility. Unlike prior studies based on lab simulations or controlled testbeds, this research is rooted in an active offshore environment where system performance directly impacts business continuity.}

\Ahmad{To address this, a full-stack platform was designed and built using modern cloud-native engineering principles. The architecture integrates DevOps automation, Kubernetes orchestration, and scalable telemetry pipelines to support AI applications in production-grade settings. The platform supports data collection from both disaggregated O-RAN nodes and vessel-mounted customer equipment, enabling a unified view of network and user conditions. This dual-layer telemetry system is a rare feature in O-RAN AI literature and provides a foundation for context-aware diagnostics.}

\Ahmad{AI was added through a tightly integrated LSTM model pipeline trained to detect temporal anomalies in real time. The model’s ability to capture sequential patterns made it suitable for recognizing subtle faults in network behavior. Its integration with the platform enables anomaly detection and automated remediation within seconds, completing a closed-loop operational cycle that remains uncommon in O-RAN practice.}

\Ahmad{The platform was deployed and validated in a real offshore environment, achieving a reduction of over 90\% in connectivity issues. This operational result demonstrates that AI can deliver a measurable impact in production disaggregated networks.}

\Ahmad{This work stands out not only for its use case and results, but also for delivering a clear architectural framework that supports standardization in AI deployment for O-RAN. It provides a blueprint for integrating automation, AI, and data engineering into complex, multi-vendor environments, where real-time performance and cross-layer visibility are essential.}

\Ahmad{By addressing both business and engineering challenges in a complete, deployable system, this research contributes a rare and practical advance in the evolving field of intelligent RANs.}

\section{Problem statement}
\label{sec:problem_statement}
\Ahmad{This section starts by describing the real business problem that led to this work. We then explain why AI was seen as a good fit for solving problems in the context of O-RAN. After that, we discuss the main challenges we faced when trying to apply AI in such a complex, multi-vendor network. These points help show the necessity of building a dedicated data Analytics platform.}
\subsection{Business problem}
\begin{figure}[h]
    \centering
    \includegraphics[width=0.48\textwidth]{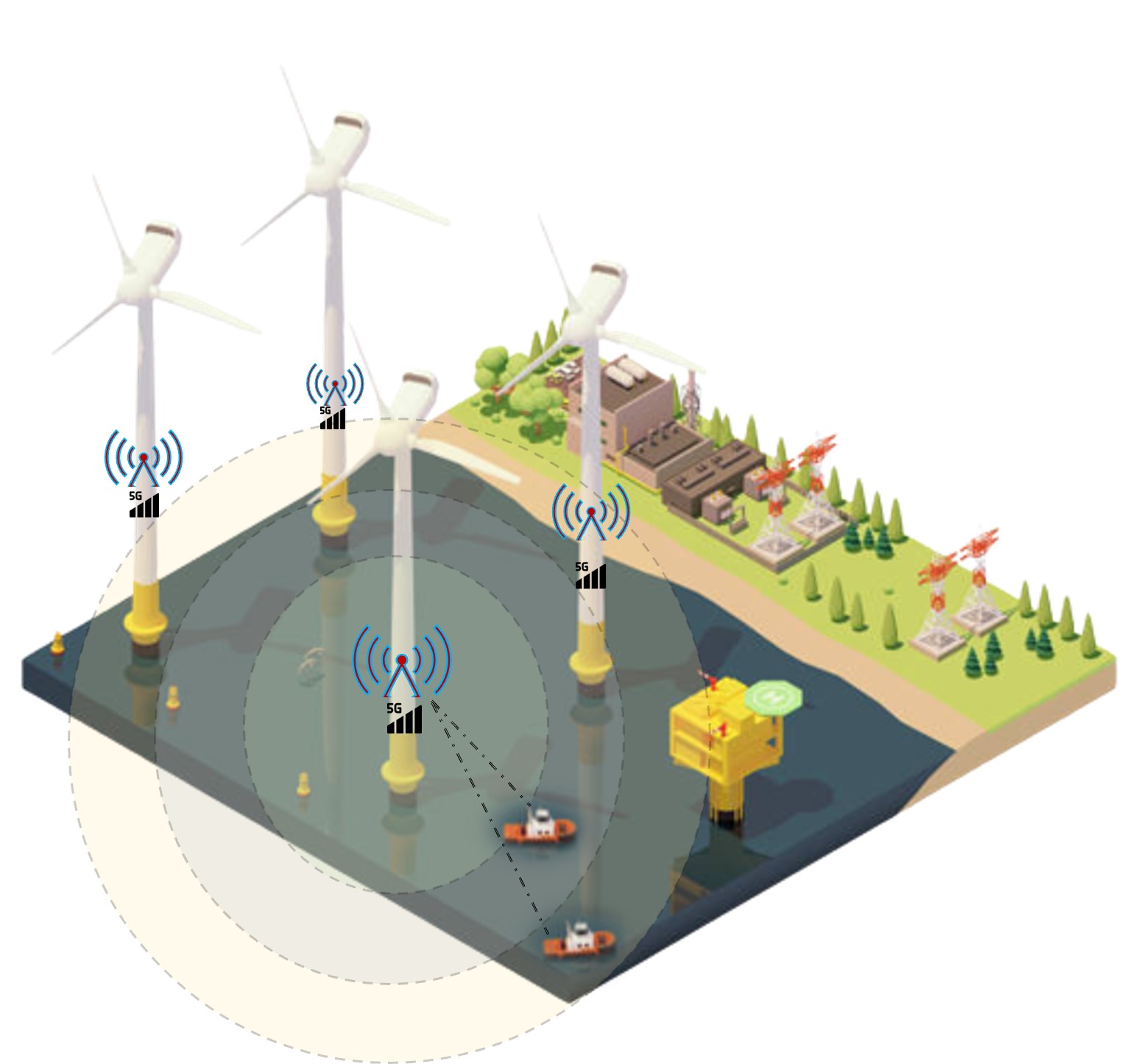}
    \caption{Illustration for the real-time offshore 5G network deployment using O-RAN and challenges in managing coverage in the sea.}
    \label{fig:sea_shore}
\end{figure}

Fig.~\ref{fig:sea_shore} illustrates the business case of deploying a commercial O-RAN-based mobile private network (MPN) in an offshore location, serving the UEs that are located in vessels that navigate around the windfarm attending to wind turbines for infrastructure operational and maintenance activities. These vessels spend most of the time navigating close to the windfarm and carry tens of people who work at sea for periods of 15 days. Therefore, these people rely on this connectivity to do their work, to communicate with their colleagues working onshore, their families, and for their entertainment. The connectivity provided by this network also supports business and operational critical processes to the organizations responsible for operating and maintaining the network, across an area of around $300\,\text{km}^2$.
Many factors such as weather conditions, sea conditions, distance to the site, and MPN operational faults may affect the quality of the connectivity service as experienced by the end-users, which might impact the: ability of running business and operational critical processes work according to the requirements or the ability of people working productively in their floating offices. Despite these challenges, the MPN’s network operator is responsible for managing, operating, and maintaining the network’s performance as per the contracted service level agreement. 

The problem addressed by the solution described in this paper occurred in an MPN service deployed by Boldyn Networks in a windfarm located in the North Sea. This network is composed of 3-cellular sites of 3 sectors each and one cellular carrier per sector, totaling 9 macro cells that cover the whole extension of the windfarm. The cellular sites are installed across the windfarm in three different turbines, where commercial-off-the-shelf (COTS) servers are used to run the containerized O-DU, and three O-RUs are installed to implement a sector each. The O-DUs deployed offshore connect, via dark-fibre light-up using long-haul SFPs, to an infrastructure of COTS servers that run the O-CU container.
On the other hand, each relevant vessel’s network is composed of a Wi-Fi network that provides IP connectivity across the vessel. This Wi-Fi network connects to Boldyn’s customer CPE that acts like an LTE broadband router back-hauling the vessels’ IP traffic to connect to the internet and the MPN customer’s enterprise network. This CPE is installed inside the vessel, and it’s connected to external 4x4 multiple-input-multiple-output (MIMO) antennae that increase the coverage and capacity of the network. Each CPE is dual-modem capable for connection resilience and traffic load balancing reasons and one MPN SIM is inserted in each modem.
The problem resides in the connection recovery procedure implemented by the CPE when one of the modems drops its link to the network. This procedure takes 5 minutes to re-establish a connection to the macro network, which is a long period of time and in some conditions increases the likelihood of disconnections occurring in both modems at the same time. In studying the problem, it was found the following phenomena:

\begin{itemize}
    \item These disconnections weren’t always correlated with the performance of the network, the radio link conditions, or the distance to the site.
    \item It could be observed drift in the behavior of radio link performance, between the two modems of the CPE, verified by monitoring indicators such as latency and throughput, even if most of the time these were connected to the same cell and benefiting from similar radio conditions.
    \item  Updating the configuration of the modem on the CPE via its API, forced the modem connection to restart instantly and established a new radio link improving the performance of the modem in all the indicators.
    \item The process of this type of restart in the modem is much quicker to implement, when the model is online, than the process of re-connection in case of radio link drop, and has the additional advantage that can be done pre-emptively whilst the other modem is showing degrading performance.
    \item The automation of this process of analysis and decision-making could significantly reduce the number of disconnections, thus having a great impact on the quality of experience (QoE) of the users being served by the network system. 
\end{itemize}

The automation of this process enabled us to develop a self-healing mechanism that implements advanced techniques. We tried conventional methods like rule-based systems, which rely on observed situations and the experience of the subject matter expert (SME). However, these methods often fall short in detecting disconnections before they occur and tend to perform many unnecessary actions, making them difficult to utilize and ineffective in handling complex and unexpected cases. In this case, using AI is considered the best approach to solve the problem, as it learns from patterns in historical data over time. For the detection of this type of performance anomaly in the modem, the model was trained on available historical performance data merged with records showing the timestamps when engineers took action to resolve the issue. This resulted in a labeled dataset that highlights the relationship between the behavior patterns of relevant performance indicators and the need for decision-making action. Some of these performance indicators included reference signal received power (RSRP), reference signal received quality (RSRQ), IP packet data latency, timestamp, location, the cell to which the modem is connected, and others from various parts of the network.

\Ahmad{The process of collecting data from the CPE device via the API every 5 seconds, preparing and enriching it, and streaming it to the proposed data platform was designed to enable real-time decision-making and direct provisioning through machine learning. In this offshore environment, where disruptions must be detected and resolved within seconds, AI plays a central role in maintaining service continuity and operational efficiency. However, this intelligence can only operate effectively when supported by a robust data platform. The platform provides the necessary infrastructure to manage, transform, and deliver high-frequency telemetry data to the model in a reliable and scalable manner. Without it, the AI pipeline would lack the consistency, speed, and data quality required for low-latency inference and autonomous action. Thus, the data management platform is not just a support layer but a critical enabler of the AI-driven solution.}

\subsection{The motivation of deploying AI in O-RAN}
AI is becoming increasingly important in O-RAN compared to traditional RAN due to its capability to address complex network demands and enhance overall performance. Several key aspects benefit from the integration of AI~\cite{polese2023understanding,recentadvancement23218792}:

\begin{itemize}
\item {\textbf{Reducing Complexity:}} O-RAN networks have a more complex, disaggregated architecture compared to traditional RAN, making manual management and optimization more challenging. Also, building traditional applications that process the data will be challenging. AI algorithms can automate and optimize these processes, like the discussed auto-healing process in this paper. It also compensates for the shortage of skilled engineers to manage such novel networks ~\cite{recentadvancement23218792}.

\item {\textbf{Real-Time capability:}} The new O-RAN architecture supports Real-time and near real-time RIC allowing AI algorithms to respond to network changes in real-time, allowing for more efficient and effective management of the network, like traffic steering~\cite{erdol2022federated}.

\item {\textbf{Cross-Layer Optimization:}} The intelligence executed in O-RAN is expected to perform cross-layer optimization over the network, which outperforms the classical optimization focusing on solely communication blocks.

\item {\textbf{Future Possibilities:}} The programmability, openness, and disaggregation enable opportunities for innovation especially when utilizing AI. One of the most important ideas that will change the shape of networks is autonomous management which will provide advanced capabilities compared to the traditional methods. Some of these capabilities are as follows:
\begin{itemize}
\item {\textbf{Improved Network Performance:}} AI-based algorithms can be used to optimize network performance by dynamically allocating network resources and adjusting network parameters based on real-time network conditions such as autonomous QoE and QoS resource optimization~\cite{polese2023understanding, recentadvancement23218792,li2024netmind}.

\item {\textbf{Cost Savings:}} By fitting autonomous network management in O-RAN, this will reduce the need for human intervention to manage the complex networks, being an essential concept to scale up the number of O-RAN networks and their size. It will also provide additional means to reduce cost by deploying specific AI applications such as automatic energy-saving applications and efficient resource utilization ~\cite{ANS10335921}.

\item {\textbf{Energy Efficiency Improvement:}} The AI capability embedded in O-RAN will be assisting in reducing the overall operational energy consumption of the O-RAN system. Apart from the software design of optimizing the network elements and control signal configuration, with AI, more flexible and fine-grained network function operation methods can be supported, for instance, the toggling off and on of carriers and cells in O-RAN can be conducted in the RIC with a non-real-time fashion~\cite{kundu2024towards}. 
\end{itemize}
\end{itemize}

In summary, the use of AI in O-RAN allows for more powerful automation, optimization, and insights compared to traditional RAN, making it a key enabler for the development and growth of the O-RAN market~\cite{BONATI2023109502, ric, AI-enabled_O-RAN, Soltani9881863,li2022transmit}.

\subsection{The challenges in enabling AI in O-RAN network}
\label{sec:AI challenges in O-RAN}
In this publication, we also focus on the engineering challenges, given our background in engineering. While there are also legal, regulatory, and business-related challenges, our work mainly deals with the technical side. As we moved forward with implementing AI in O-RAN, we faced several challenges as follows:

\begin{itemize}
\item {\textbf{Multi-vendor RAN Model}}: The multi-vendor model involves deploying and operating RAN equipment and software from different vendors within the same network. This raises significant challenges, especially in centralizing management and collecting data from these systems. Data integration becomes more difficult, delaying the development of AI applications. Additionally, having multiple O-RAN models can isolate each component, limiting the data available for building holistic applications. Consequently, many existing platforms don't meet the unique requirements of multi-vendor O-RAN networks, slowing AI model development and complicating the integration of developed models within other O-RAN systems deploying different vendors. 

\item {\textbf{Big Data Characteristics}}: O-RAN networks generate highly complex, diverse, and high-volume data, such as network performance data, configuration management data, fault management data, infrastructure data, and user equipment trace data. The big data characteristics of these data sources pose challenges in storing, processing, and analysing information. Moreover, the varied structures and formats of data make integration into existing analytics platforms difficult. 

\item {\textbf{Integration with Existing Systems}}: Building certain AI models with specific algorithms requires data from external sources not directly accessible to the O-RAN platform, such as UE and network functions of other domains, such as transmission networks or core networks. This external data is essential for comprehensive analysis and characterization of performance across the overall network ecosystem, but poses integration challenges.

\item {\textbf{Standardization}}: The lack of standardization in AI for O-RAN and RIC APIs presents significant challenges. There is no guarantee that developed xApps or rApps will be reusable across different RICs. Additionally, the industry is still in the early stages of establishing a centralized RIC system for the multi-vendor O-RAN model, further complicating standardization efforts.

\item {\textbf{Skilled Resources}}: Managing AI algorithms in O-RAN networks requires specialized skills and expertise, which may not be readily available in the market. A deep understanding of O-RAN is essential for those involved in AI development for O-RAN, making it difficult to find qualified personnel.

\end{itemize}

The primary challenge lies in the absence of suitable data analytics platforms capable of accommodating the distinct data and AI needs in multiple O-RAN networks built using multiple O-RAN vendors. To overcome this obstacle, it is imperative to develop new data analytics platforms specifically designed to meet the unique requirements of O-RAN networks. The platform must seamlessly integrate with existing O-RAN network systems and associated infrastructure components. They should have the ability to process and analyze vast quantities of complex and varied data and deliver use cases that depend both on real-time (or near-real-time) capabilities, whilst facilitating the integration of data from other non-O-RAN diverse sources.

Given that the approach to O-RAN is still an evolving technological concept, there is significant potential to contribute to the development of more robust systems. These challenges have driven us to create a unique cloud-native data analytics platform. Our goal is to directly address these difficulties and provide an environment that supports the development and execution of AI applications within the O-RAN ecosystem.

\section{Proposed cloud-native open data platform for multi-Vendor O-RAN}
\label{sec:data_platform}
Before introducing our cloud-native open data platform, we first outline the key engineering challenges associated with managing data across diverse O-RAN network instances. We also highlight the critical need for a standardized, scalable AI development workflow capable of operating across multi-vendor O-RAN environments. 
\subsection{Data management challenges in multi-vendor O-RAN deployments}
Fig.~\ref{fig:ORAN_architecture} depicts a real-world scenario in which a communications service provider (CSP) deploys multiple O-RAN network instances to support an ecosystem of MPNs and neutral host networks (NHNs).

Vendor A has been selected to deliver the MPNs; Vendor B supports outdoor high-density demand (HDD) areas through a multi-operator RAN (MO-RAN) design; and Vendor C provides in-building MO-RAN services. In both MO-RAN use cases, each mobile network operator (MNO) runs on its own virtualized instance. These choices reflect the architectural and functional needs of specific deployments.

These networks rely on different vendor platforms and are deployed in various locations, such as airports, offshore windfarms, smart cities, stadiums, and hospitals. Each deployment varies in vendor selection, internal architecture, supported MNOs, and feature sets.
There are other network components, such as the IP network in dark blue, and IT \& Security in yellow (which is shared by all vendors). Other network components are also used like routers, mobile apps, and IoT devices, each of which supports the operation of the network.

The authors of this paper have collaborated closely to address the challenges involved in developing AI/ML models within such heterogeneous O-RAN ecosystems. A significant gap has been identified: \textit{there is a lack of a supportive platform for managing data and running analytics across multi-vendor O-RAN systems, as well as for deploying AI models within them.} This gap highlights the need for a unified approach to data integration and AI enablement in operational networks.

Accordingly, we designed and implemented a multi-vendor, cloud-native open data architecture that supports AI and Data Analytics. The platform centralizes and standardizes the management of multi-vendor networks by integrating data sources across subsystems into a unified overlay. This enables simplified access and streamlined analytics through standardized data pipelines, avoiding fragmented, case-by-case integration.

This approach introduces a role-based model for the whole AI application development:
\begin{itemize}
    \item The subject matter expert (SME) defines the use case, data structure, and analytical goals, as well as the criteria for validating outcomes.
    \item The data engineer builds the required data pipelines in collaboration with the SME and the data scientist. He also applies CI/CD and MLOps practices to the data and Apps processes.
    \item The data scientist to do data analytics and train, test, and validate AI/ML models that meet the use case objectives.
\end{itemize}

In operational O-RAN environments, acquiring, storing, and processing data for AI model training is a significant challenge. Although standard interfaces like E2 offer access to network components such as the O-DU and O-CU, the extracted data is typically raw and lacks schema standardization, making it unsuitable for direct use in AI pipelines. In order to achieve this, we introduce a structured multi-stage process that reflects how effort is distributed across the AI lifecycle within O-RAN environments in the next section.

\subsection{The AI development pyramid in O-RAN}
\label{subsec:AI_Pyramid}

To effectively leverage AI within O-RAN and its interfaces, a multi-stage process is necessary. Initially, raw data from various sources must be collected, validated, enriched, transformed, and consolidated into an integrated data pool. This prepares the data for processing by using data engineering techniques, such as applying business rules, calculating key performance indicators (KPIs), performing feature engineering, and linking data tables based on network topology mapping. These processes ultimately enable the application of algorithms tailored to specific use cases.

Moreover, an O-RAN network is constructed on top of other system components, including IP networks and private cloud server infrastructure. The operation and maintenance of these components are vital for overall network performance and should be seamlessly integrated into a holistic network management process that encompasses all system elements.

\begin{figure*}[!h]
    \centering
    \includegraphics[width=0.9\textwidth]{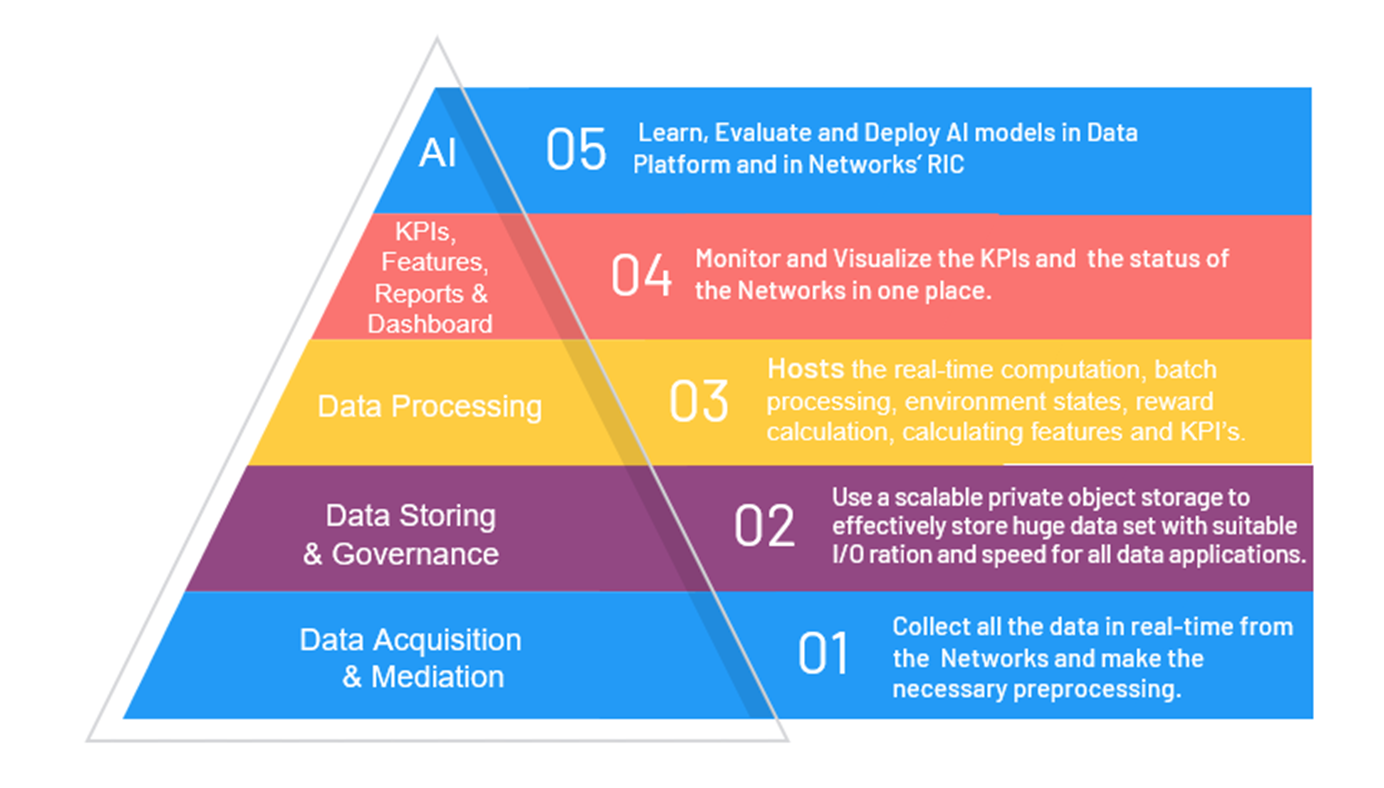}
    \caption{Cloud-native Data platform workflow with respect to the applied efforts.}
    \label{fig:Work-efforts-triangle}
\end{figure*}

The primary goal of this platform is to streamline and accelerate the development and hosting of AI solutions within the O-RAN ecosystem. As shown in Fig.~\ref{fig:Work-efforts-triangle}, the largest share of development effort lies at the foundation: the data acquisition and mediation phase. It is during this foundational phase that the collaborative work between the three roles, mentioned earlier, is most intensive. When executed properly, the components developed at this level support and simplify the work in subsequent phases, reducing the effort required at each step. Ultimately, the AI model deployment phase benefits from having all required components readily available from the outset.

This efficiency stems from eliminating redundant work across multiple network deployments. As data is centralized and handled through a unified system, the handling process becomes standardized, allowing the same AI models to be adopted across different systems from various vendors.

The subsequent phases of the AI lifecycle include data storage and processing. While these require comparatively less effort, they are essential to enhance the performance of the platform and to prepare datasets in line with the SME’s analytical requirements. At the top of the pyramid lies the development of AI models, an activity made more effective and manageable by the availability of high-quality, well-organized data.

For example, this structured approach ensures that developing AI models becomes more straightforward and reliable, supported by a standardized foundation. It also facilitates the reuse of components, enabling faster integration of new network instances and easier adaptation of AI models across deployments. To further illustrate this structured AI development process, the next section presents the platform’s layered architecture in detail and its role in enabling scalable, multi-vendor O-RAN operations.

\subsection{The proposed platform architecture layers}

\begin{figure*}[!h]
    \centering
    \includegraphics[width=1\textwidth]{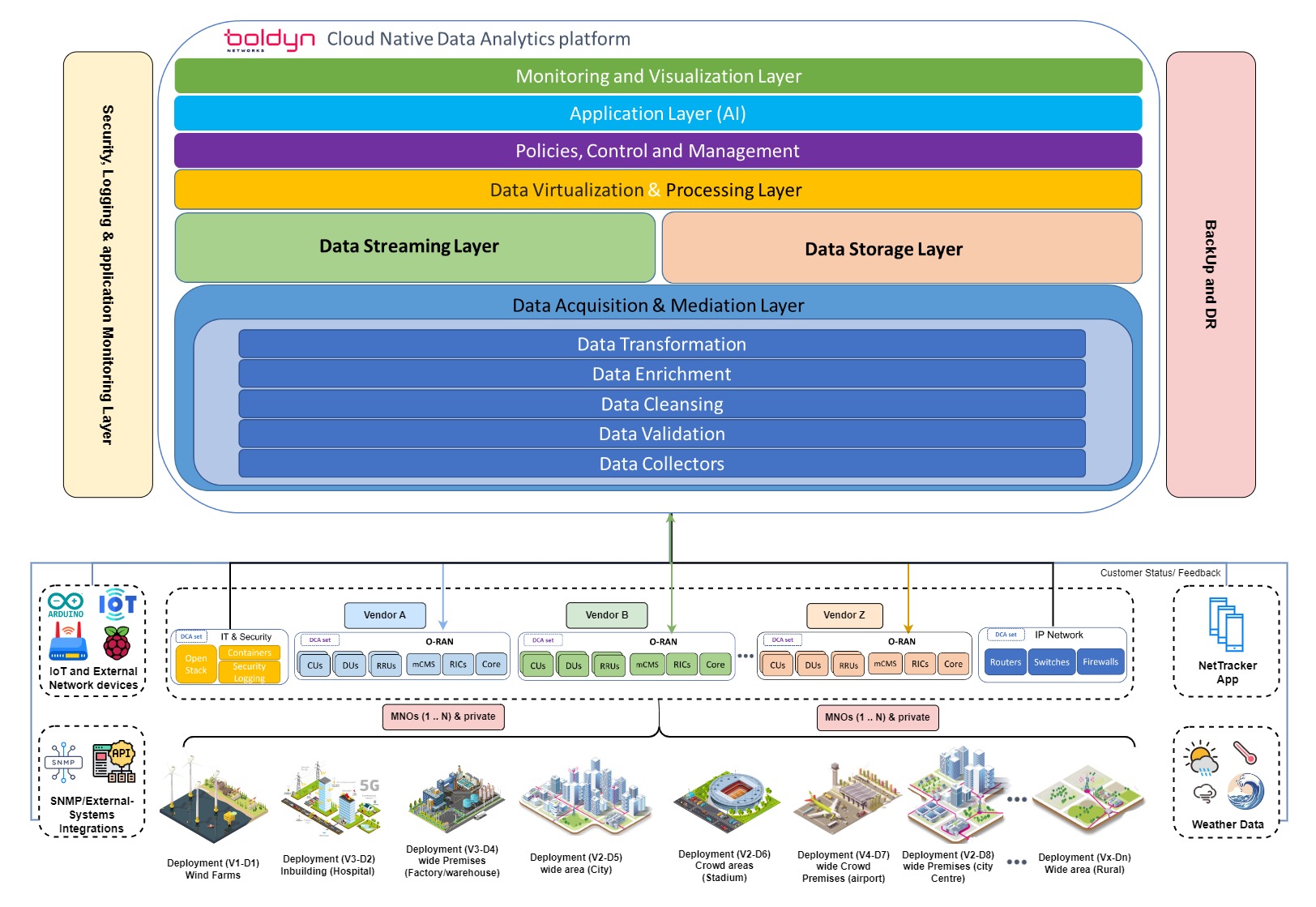}
    \caption{Boldyn Networks Cloud native Data Analytics platform architecture layers.}
    \label{fig:data-platform}
\end{figure*}

Fig.~\ref{fig:data-platform} depicts the main layers and components of the data management system and its interfaces with network functions and management systems. The data mediation layer includes tools to connect with multiple data sources, collect the data, and perform initial processing to validate, enhance, and unify the datasets. This creates a coherent data stream from multiple instances of the same data source type.

The datasets are then stored in the data storage layer or streamed directly to upper layers, including the Data Virtualisation \& Processing Layer, Application Layer, or Data Visualisation and Monitoring Layer. When the data is cleansed, validated, and enriched, it is processed in the processing layer using big data execution engines or virtualization techniques, depending on the AI use case.

The Policies, Control, and Management Layer contains information about the network topology, data mapping, and business roles. These are applied during processing to produce richer datasets, features, and KPIs that are used in the AI Layer and/or the Visualisation and Monitoring Layer, which acts as the single pane of glass.

In the Visualisation Layer, network engineers can access reports and dashboards that combine datasets from multiple sources. These provide insights into network performance, configuration, and faults—supporting analytical use cases such as service assurance and operational situational awareness.

Despite the layered depiction, the order or positioning of the layers does not necessarily indicate a hierarchy or sequence. Each layer can be accessed directly, independent of its position in the stack. The layers are broken down as follows:

\subsubsection{Data collection agent (DCA)}
A DCA is a self-built software application deployed across network equipment. It is developed to extract or generate data from a function or interface that is not readily available and is considered important for implementing one or more use cases.

\subsubsection{Data acquisition and mediation layer}
This layer handles the core integration tasks for all deployed networks and supporting systems. Data is collected from network functions and devices, including O-RAN, IoT devices, UEs, customer premise equipment (CPE), and others. It processes data in various formats and structures such as CSV, JSON, XML, unstructured or semi-structured text, APIs, SFTP, and SNMP.

The goal is to unify how data is made available to subsequent layers by transforming and enhancing content. This includes data parsing, enrichment with schema or metadata, format transformation, and distribution to upper layers. This layer plays a key role in standardizing data access across the platform.

\subsubsection{Data storage layer}
This layer stores the collected data. Depending on volume and access needs, suitable storage systems are used, such as data lakehouse technologies for large datasets, or relational databases for smaller sets and mapping information.

\subsubsection{Data streaming layer}
This layer delivers data to the application or processing layer in real time, avoiding delays from traditional storage-and-retrieval cycles. It is essential for use cases involving online monitoring or real-time decisions. A parallel storage process runs independently to preserve this data without affecting performance.

In this work, it was implemented for an anomaly detection AI/ML model that requires continuous, low-latency data streams. This model processes data before storage, which is critical for near-RT RIC and RT-RIC scenarios where decisions must be made within milliseconds.

\subsubsection{Data virtualization and processing layer}
As data reaches the storage or streaming layer, it is not always immediately usable. Two technologies are applied here: big data execution engines and data virtualization.

Execution engines such as Spark perform intensive calculations on large datasets, such as generating KPIs and performing feature engineering. Data virtualization enables SQL-like access by joining data from lakehouse systems, relational databases, and other sources. For instance, calculated KPIs can be linked with network topology and other data to build views for upper layers.

\subsubsection{Policies, controls and management}
This layer enforces business rules and defines relationships across the data layers and the AI application layer. It manages O-RAN FCAPS~\cite{ITU2024} metadata, network topology, alarms, policies, API access, and network change activities.

It also acts as a central mapping layer, standardizing data from different vendors into a common format. This simplifies cross-network operations and improves overall data coherence.

\subsubsection{Application layer}
This is where the lifecycle of data-driven applications is finalized. For AI, it is the environment where SMEs and data scientists train, test, publish, and validate models.

\subsubsection{Data visualisation layer}
This layer implements business intelligence capabilities, offering SMEs dashboards and reports that combine data from various sources and stages. It enables system-wide performance monitoring and operational awareness. It also acts as the interface between the user and the AI models by reporting AI actions during routine operations. More details about this layer are available in~\cite{Peizheng9931127}.

Together, these layers form a flexible and scalable architecture that supports real-time analytics and the integration of AI applications across multi-vendor O-RAN environments.

\subsection{Platform core components}

\begin{figure*}[!h]
    \centering
    \includegraphics[width=0.95\linewidth]{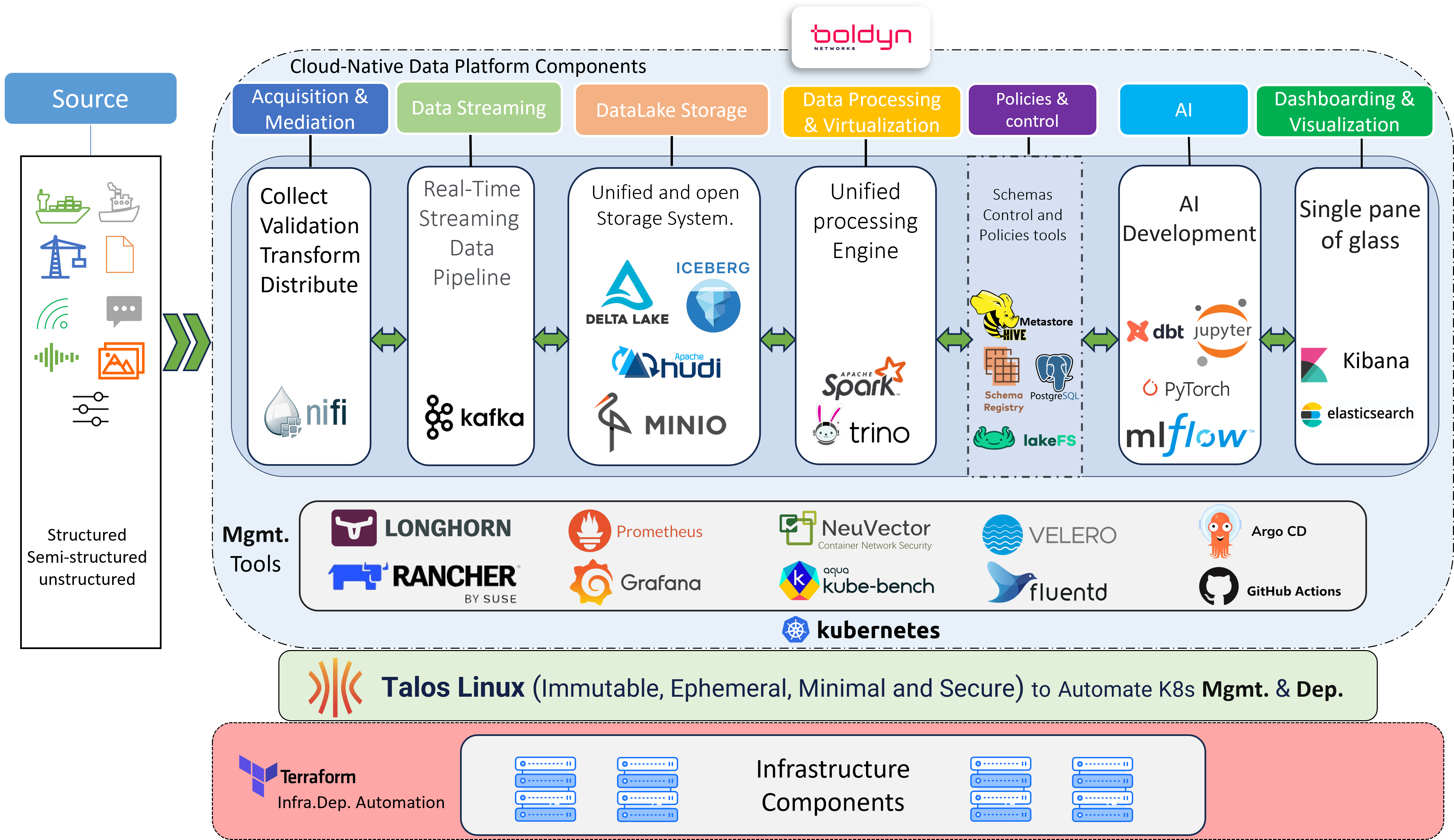}
    \caption{Boldyn Networks cloud-native data analytics core components. This figure shows the main components related to infrastructure, operating systems, Kubernetes management, and data core components.}
    \label{fig:components_of_data_platform}
\end{figure*}

Fig.~\ref{fig:components_of_data_platform} depicts the architectural building blocks of the cloud-native data management platform. This platform is further detailed as follows.

The first tier, following a bottom-up order, contains the infrastructure and its automation toolset. We utilized Terraform to define and provision our infrastructure as code, enabling efficient management and automation of the infrastructure resources that run the operating system (the second tier of the architecture), Talos OS. Talos OS is a modern, Linux-based operating system specifically designed for Kubernetes. It provides a secure, minimal, and immutable platform to enhance security. One of its major features is that it automates and simplifies the management and operation of Kubernetes clusters. On top of this, Kubernetes is installed to deploy the microservices and applications that implement both cluster management functions and the data management application system (this is the third tier).

The following applications and services are implemented in the third tier:

\subsubsection{Data management applicational systems}
\begin{itemize}
    \item Apache NiFi is the core of the acquisition and mediation layer. It is used to automate and manage the flow of data between systems where the data sources are located (across multiple platforms and their components) and other applications and systems of the data management platform that consume this data. Therefore, it is the main application used to implement the collection and mediation layer of the platform. 
    \item Apache Kafka is the core streaming system and is used to transfer data in real time from the source to the processing layer.
    \item In the storage layer, we use the object storage system MinIO, which provides an S3-like API. Depending on the data use case, we store the data using one of several techniques such as Hudi, Delta Lake, or Iceberg open tables, or even in its raw form.
    \item The data is later accessed by Apache Spark to perform complex big data processing on data from streaming, data lakehouse, databases, or from all sources combined. Trino, on the other hand, is a virtualization system used to perform SQL-like queries on datasets available throughout the platform.
    \item In the application layer, we use Python, Jupyter Notebook, and MLflow to train, test, and validate the AI modules before publishing their microservice deployments in the processing layer. We also use DBT for version-controlled analytics workflows.
    \item In the control and management phase, we use Apicorio as a schema registry, Hive Metastore to store data catalogues and metadata, and PostgresDB to store the network topology, rules, and alarm triggers.
    \item The final stage of the data process involves Elasticsearch and Kibana, which are used to visualize data from multiple sources, including applied AI actions and outcomes, in a unified dashboard that serves as the platform’s single pane of glass.
\end{itemize}

\subsubsection{Kubernetes cluster management tools}
There are other systems included in the management layer. These tools support the operation of the data components across the various layers:
\begin{itemize}
    \item Longhorn \& Rancher: for Kubernetes and cloud-native storage (data layer for applications) management. 
    \item Prometheus and Grafana: for platform monitoring and alerting. 
    \item NeuVector and Kube-bench: for security and vulnerability checks.
    \item Velero and Fluentd: for log and backup management.
    \item Argo CD and GitHub Actions: for CI/CD and GitOps.
\end{itemize}

\subsubsection{CI/CD pipeline}
The development of data pipelines on this architecture can be complex, tedious, and error-prone if done manually. A set of tools and systems is used to automate the development and deployment of these pipelines. GitOps methods, such as CI/CD, simplify, automate, and manage this process. 

\begin{figure}
    \centering
    \includegraphics[width=0.9\linewidth]{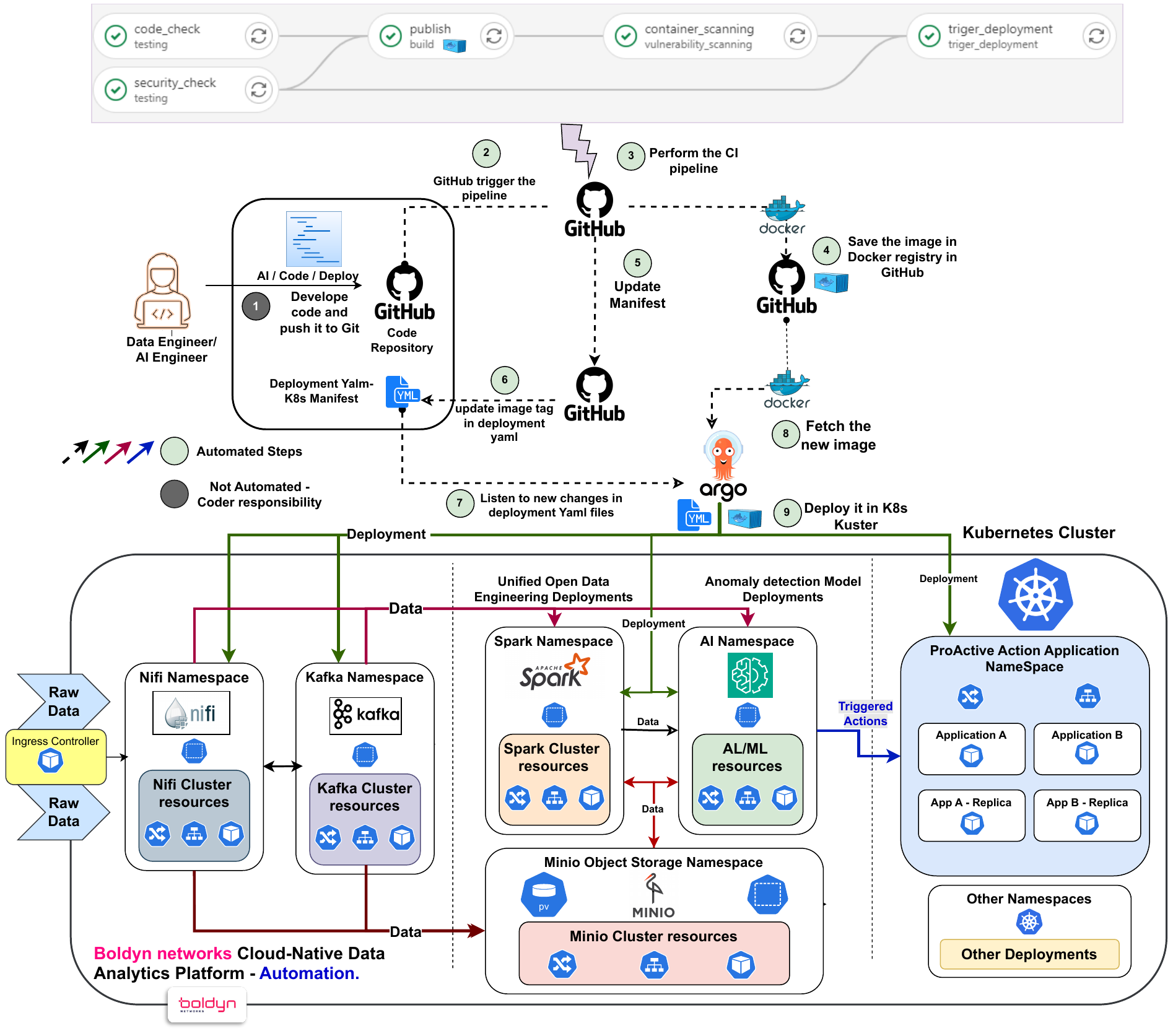}
    \caption{\textbf{ End-to-end automation pipelines for AI and data engineering processes using GitOps and CI/CD practices.} \Ahmad{This figure illustrates the fully automated lifecycle of developing, testing, containerizing, and deploying data pipelines and AI/ML models within our Kubernetes-based platform. Leveraging GitOps principles, the process begins with a data or AI engineer committing source code and deployment manifests to GitHub. This triggers a CI/CD pipeline that conducts code quality checks, security scans, and builds Docker images. These images are stored in a secure registry, and deployment manifests are automatically updated with the latest image tags. Continuous Deployment (CD) is handled by ArgoCD, which listens for manifest changes and deploys updated applications to the appropriate Kubernetes namespaces. The platform integrates raw data ingestion (via NiFi and Kafka), distributed processing (via Apache Spark),  AI model deployment and other applications deployments into a seamless workflow. This architecture ensures Automation, simplicity, security, scalability, and reliability for production-grade AI and data analytics applications.}}
    \label{fig:AI-application-automation}
\end{figure}

Fig.~\ref{fig:AI-application-automation} shows an example of a CI/CD process implemented to develop and deploy the data pipeline of one use case addressed by the platform. It provides details about each task involved in data engineering and AI model preparation.

The pipeline CI/CD cycle begins by developing the source code and publishing it to the GitHub repository, which triggers a set of automated workflows to check the quality of the code and perform security scans. In the next step, the code is built into Docker images through a process known as containerization. All Docker images are then scanned for vulnerabilities before being stored in the Docker registry. The final step of the CI process is to update the deployment manifests with the new Docker image. The CD process, handled by ArgoCD, automates fetching the latest manifest changes and deploying the updated application.

\subsection{Features of this platform}
\label{subsec:feature of the platform}
The platform is designed to offer flexibility in working with O-RAN data and includes the following features and benefits:

\begin{itemize}
\item \textbf{Unification}: The platform can collect, store, and process all types of data from various sources, regardless of volume, velocity, and variety, due to its modern open data architecture. It enables batch and real-time processing, allows analysts to perform advanced data exploration, supports the development and deployment of AI models including federated learning, and provides comprehensive tools for reporting and interactive visualization across complex network environments.

\item \textbf{Cost-effective}: Built on commodity servers, the platform leverages Kubernetes (K8s) container orchestration and other tools to simplify management and reduce the required resources. In addition, using open data architecture and data lakehouse technologies helps consolidate data, optimize query performance, reduce bandwidth usage, and leverage open-source tools, improving collaboration and overall usability.
 
\item \textbf{Automation and Standardization}: Using DevOps and GitOps~\cite{GitLabGitOps2024} technologies, all infrastructure, setup, data, and AI pipelines are fully automated. This reduces the operational workload for development and deployment, making the development cycle predictable. Business owners can collaborate more effectively with developers to create solutions, ultimately reducing the resources needed.
\item \textbf{Scalability}: As more O-RAN networks are deployed, the ability to scale becomes critical. The platform has been thoroughly evaluated to ensure it can handle growing demands in AI workloads, data processing, and storage capacity.

\end{itemize}

\section{The proposed solution}
\label{sec:proposed_solution}
\PL{For the specific business problem introduced in Sec.~\ref{sec:problem_statement}, this section will discuss the anomaly types and model selection, data sources, LSTM model design, model deployment procedure, and validation results.}

\subsection{Anomaly types and model selection}
Network anomaly detection or prediction is a complicated task. Anomalies are referred to as patterns in data that do not conform to a well-defined characteristic of normal patterns. Anomalies can be classified into three types: (1) point anomaly can be considered as a particular data instance deviation from the normal pattern of the dataset; (2) contextual anomaly is defined when a data instance behaves anomalously in a particular context; (3) collective anomaly happens when a collection of similar data instances behave anomalously with respect to the entire dataset, the group of data instances is termed a collective anomaly~\cite{AHMED201619}. Existing anomaly detection methods include NN, support vector machine or rule-based classification, statistical signal process method and clustering.
\Ahmad{In the case of the offshore O-RAN deployment studied in this work, the prediction task involves detecting abnormal behavior in modem connectivity based on historical telemetry and operational data. These anomalies are best characterized as point anomalies, as they are typically reflected in sudden shifts in performance indicators such as RSRP, RSRQ, latency, and signal quality. Given the temporal nature of these patterns and the necessity of capturing sequential dependencies in the data, LSTM networks were selected as the most suitable model. LSTM architectures are well-suited for modeling time-series data with complex temporal dynamics, allowing the system to identify early signs of degradation that precede connectivity loss.
The choice of LSTM aligns with the objective of the platform to provide real-time, proactive anomaly detection within a streaming data environment. This integration enables the model to not only identify anomalous conditions but also trigger automated corrective actions with minimal latency, improving the resilience of the O-RAN deployment.}

\subsection{Data specification}
\Ahmad{The dataset used in this study was collected from a production deployment of a private offshore mobile network based on a O-RAN architecture. This network serves vessels operating within a windfarm environment and was instrumented with custom telemetry agents and monitoring tools to support continuous observation. Data was captured across both network-side components, like the O-CU, O-DU, and O-RU, and from onboard CPE installed on the vessels themselves. Together, these sources provided a comprehensive view of both infrastructure performance and user-side experience in a complex and mobile offshore setting. Network-side data was in line with 3GPP specifications. This included a range of FCAPS metrics such as latency, throughput, availability, signal quality indicators, Handover, fault and alarm records, configuration snapshots, and many other detailed data. The network-side data also contains data related to the radio coverage prediction that shows the expected coverage in this specific deployment. On the user side, CPEs installed aboard vessels transmitted telemetry that captured device-level QoS metrics (such as RSRP, RSRQ, and cell transitions), GPS-based geolocation, velocity, and diagnostics like speed test results.}

\Ahmad{The data was highly varied in structure. Some arrived in well-organized formats such as CSV files or relational tables, while other streams came in semi-structured formats like JSON or XML, or as raw, unstructured logs. These were delivered through different channels, including Kafka topics, RESTful APIs, secure file transfers, and compressed archives. The sampling frequency also varied, ranging from high-resolution, near-real-time updates at five-second intervals to lower-frequency aggregated records, depending on the source and purpose of the data.
In total, the platform ingested more than 1.5 TB of data each month, representing over 1.5 billion individual records. To properly handle this volume and variety, the system employed a set of scalable data engineering processes that included compression, data selection, enrichment, quality checks, and time alignment. These efforts were essential to ensure the consistency and usability of the data, particularly given the complexity of managing multiple vendor systems in O-RAN.}

\subsection{Data preparation and feature engineering}

\Ahmad{Before training the LSTM model, considerable attention was given to refining the dataset to include only meaningful operational conditions. One key challenge was the presence of data collected while vessels were outside the agreed service area, which is beyond the network's designed coverage. These periods, which occur regularly as part of normal vessel navigation, often produce telemetry that does not reflect an active or stable connection to the offshore network. Including such data could negatively affect the learning process by creating misleading associations between poor connectivity and expected behavior.}

\begin{figure*}   
    \subfloat[\label{fig:ue_trace}]{
      \begin{minipage}[t]{0.5\linewidth}
        \centering 
        \includegraphics[width=3.2in]{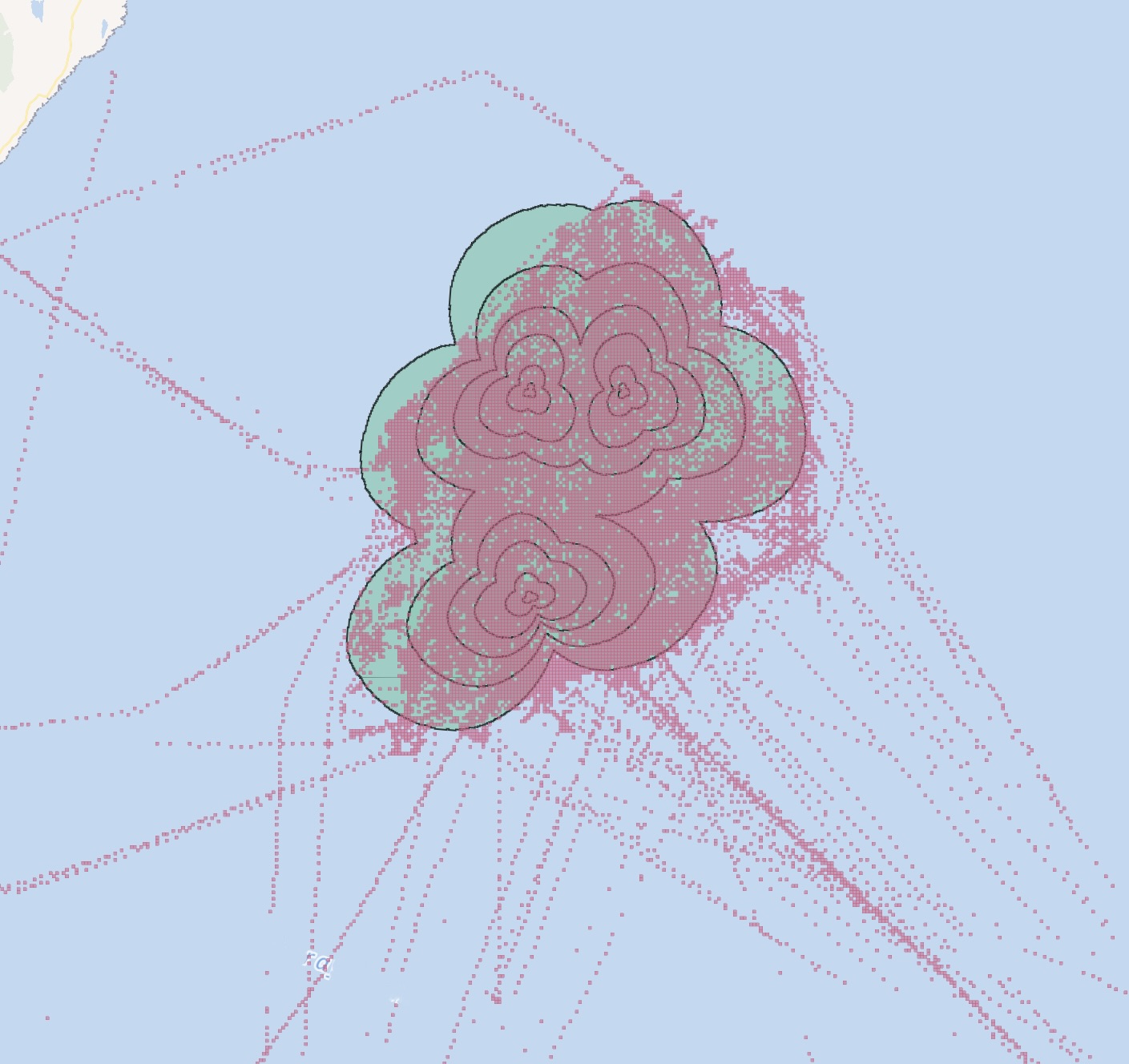}   
      \end{minipage}%
      }
        \subfloat[\label{fig:coloured_coverage}]{
      \begin{minipage}[t]{0.45\linewidth}   
        \centering   
        \includegraphics[width=3.04in]{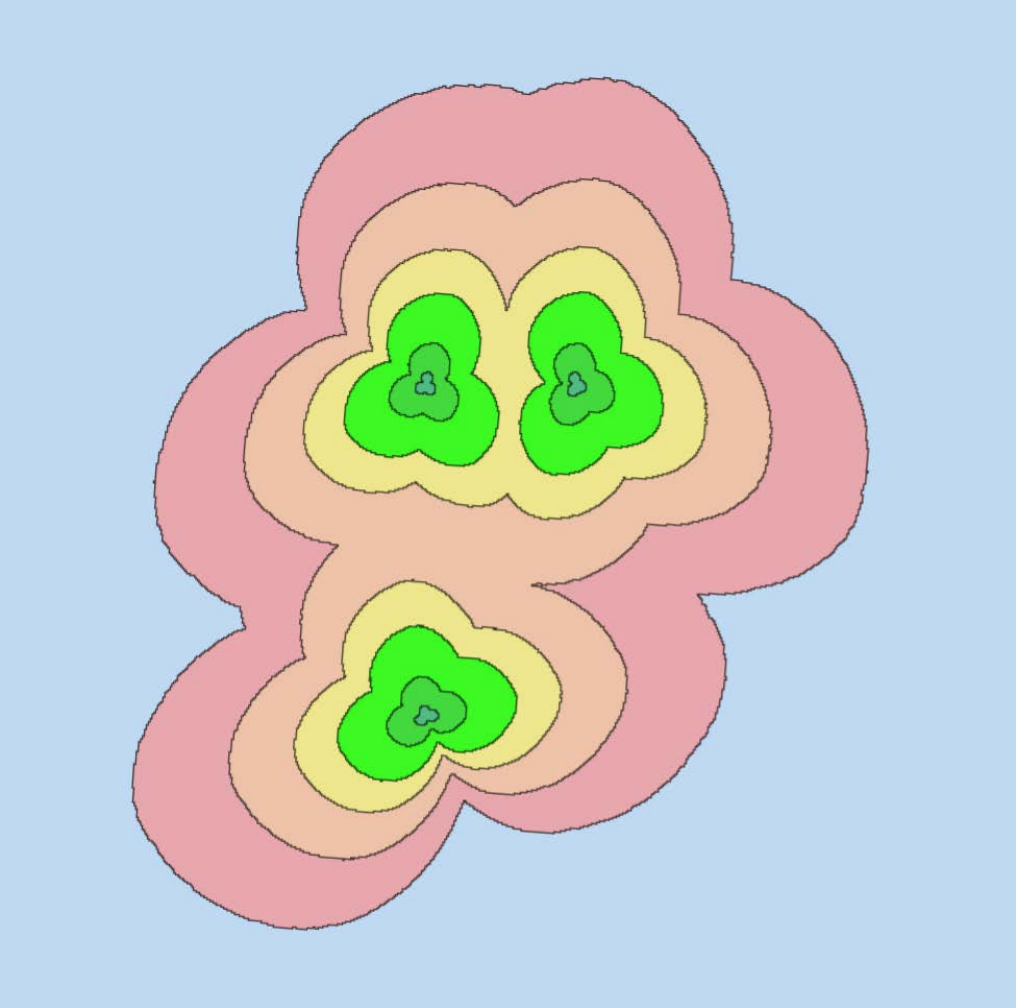}   
      \end{minipage} 
      }
      \caption{(a) Visualisation for the location of the vessel from the data used in the AI model layered on top of the predicted network coverage. (b) Predicted radio coverage map used for data filtering and feature engineering. Signal strength levels range from strong (green) to weak (darker red shades).
      } \label{fig:2figures_ue_trace_coverage}
      \vspace{-0.2cm}
\end{figure*}

\Ahmad{The dataset used in this study combines multiple sources, including network telemetry, onboard CPE data (which includes geolocation), and radio coverage predictions. Merging these sources gave us the full context needed to understand where and when each measurement was taken is relevant to the model. This integration was not only important for filtering out irrelevant records but also played a key role in data engineering. It enabled the creation of new features based on the vessel’s location, observed network performance, and expected signal levels from the coverage map. Without this combined view, building a reliable dataset for training the model would have been far more difficult and time-consuming. An example of this alignment is shown in Fig.~\ref{fig:ue_trace}, which visualizes vessel location data overlaid on the predicted network coverage map. As seen in the figure, many data points fall outside the intended coverage boundaries, demonstrating the need for location-based filtering \footnote{ The latitude and longitude values are removed for the consideration of the clients' privacy.}. To address this, we first applied a filtering step to exclude all records where location data or time period placed the vessel clearly outside the coverage boundaries.}

\Ahmad{To illustrate how features were engineered from the integrated data, we used measurements reported by the CPEs such as RSRP, RSRQ, latency, and modem status at five-second intervals, along with GPS coordinates. These were matched against a pre-modeled radio coverage layer that defines expected signal levels across the entire offshore windfarm area. As shown in Fig.~\ref{fig:coloured_coverage}, the predicted coverage map was derived from RF planning tools and segmented into levels, where Level 1 (green) indicates the highest expected signal strength, and lower levels (shaded toward red) indicate progressively weaker coverage.}

\Ahmad{A key feature extracted from this integration was a coverage performance consistency check. For each CPE measurement, the current observed RSRP was compared against the average expected RSRP for that specific geolocation. If the vessel was located in a Level 1 (green) zone, then a certain threshold of signal strength was expected. Any significant deviation from this expected performance, even when the vessel is physically inside a strong coverage area, was flagged and included as a new input feature to the anomaly detection model.}

\Ahmad{Another example, the distance between the vessel and all known cell sites was calculated. This enabled us to extract not only the distance to the currently connected serving cell but also to determine the closest neighboring cell not currently connected to the CPE. This "nearest-non-serving cell" feature is particularly relevant in assessing whether a better handover decision could have been made.}

\Ahmad{Meanwhile, the handover activity is tracked. Frequent handovers, especially when occurring in low-mobility or stationary scenarios, were considered potential indicators of instability. The frequency and duration of connections to each serving cell were included as dynamic features in the model's input sequence.}

\Ahmad{We also introduced a binary feature that indicated whether the vessel was connected to the list of expected best sites in that location (based on the coverage map) at each timestamp. This “connected-to-best-cell-in-zone” flag, combined with the performance consistency check, helped distinguish between anomalies caused by genuine radio conditions and those arising from device-side or transient network behavior.}
\Ahmad{These spatial, temporal, and performance-based features, along with many others not described here because of their large number and proprietary nature, were included in the LSTM model’s training dataset to help it detect and prevent connectivity problems before they happen\textsuperscript{\ref{datadiscloserfootnote}}. }
Table \ref{tab:engineered_features_full} presents some of the features we calculated to feed the LSTM model.

\begin{table}[ht]
\centering
\caption{\PL{Summary of some of the engineered and raw features for anomaly detection model}}
\begin{tabularx}{\linewidth}{>{\raggedright\arraybackslash}l 
                                >{\raggedright\arraybackslash}X 
                                >{\raggedright\arraybackslash}l}
\toprule
\textbf{Feature Name} & \textbf{Description} & \textbf{Type} \\
\midrule
\texttt{distance\_to\_serving\_cell} & Euclidean or Haversine distance from vessel's GPS location to the currently connected cell site. & Continuous \\
\texttt{distance\_to\_nearest\_cell} & Distance to the geographically closest available (but not connected) cell site. & Continuous \\
\texttt{rsrp\_expected\_zone} & Average RSRP value expected at the vessel’s current geolocation based on RF coverage prediction. & Continuous \\
\texttt{rsrp\_deviation\_from\_expected} & Difference between the observed RSRP and expected RSRP at the same location. & Continuous \\
\texttt{connected\_to\_best\_cell\_in\_zone} & Boolean flag indicating if the vessel located in the current location connected to any of the best cells available in that zone. & Binary (0/1) \\
\texttt{handover\_count\_window} & Number of cell handovers during a defined time window. & Integer \\
\texttt{connected\_cell\_duration} & Time duration (in seconds) connected to the current serving cell. & Continuous \\
\texttt{wan\_id} & Identifier for the wide area network interface used by the CPE. & Categorical \\
\texttt{carrier} & Mobile network operator currently serving the connection. & Categorical \\
\texttt{rsrp} & Measured  Reference Signal Received Power. & Continuous \\
\texttt{sinr} & Measured  Signal-to-Interference-plus-Noise Ratio. & Continuous \\
\texttt{rsrq} & Measured  Reference Signal Received Quality. & Continuous \\
\texttt{latency} & Round-trip latency as experienced by the CPE. & Continuous \\
\texttt{time} & Timestamp of the measurement event. & Time \\
\texttt{cell\_and\_network\_metrics ...} & Aggregated cell availability, fault indicators, and thousands of additional performance metrics collected from the RAN and core network layers. & Mixed \\
\bottomrule
\end{tabularx}
\label{tab:engineered_features_full}
\end{table}

\subsection{LSTM model design and training}

\begin{figure}[t]
    \centering
    \includegraphics[width=0.5\textwidth]{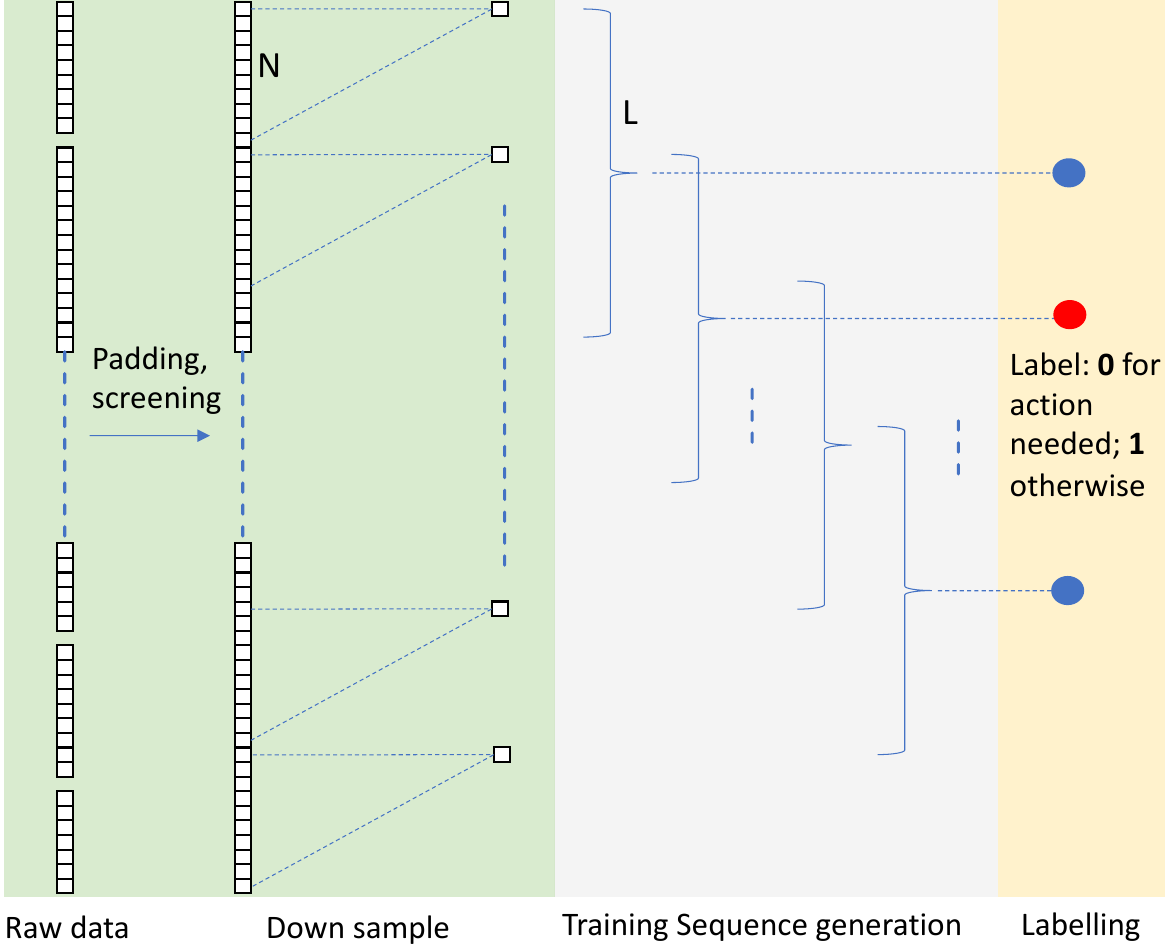}
    \caption{Data pre-processing and labelling scheme.}
    \label{fig:raw_data pre-processing}
\end{figure}

\begin{figure*}
    \centering
    \begin{tikzpicture}[node distance=2.5cm, auto]
        \tikzstyle{block} = [rectangle, draw, fill=yellow!20, 
                             text width=4em, text centered, rounded corners, minimum height=4em, line width=0.3mm]
        \tikzstyle{processing} = [rectangle, draw, fill=blue!20, 
                             text width=4em, text centered, rounded corners, minimum height=4em, line width=0.3mm]
        \tikzstyle{arrow} = [thick, ->, >=latex, line width=0.3mm]

        \node [block] (input) {Input};
        \node [block, fill=green!30, right of=input, node distance=3cm] (lstm) {LSTM};
        \node [processing, right of=lstm, node distance=3cm] (fcn1) {FCN};
        \node [processing, right of=fcn1, node distance=3cm] (fcn2) {FCN};
        \node [block, right of=fcn2, node distance=3cm] (output) {Output};

        \draw [arrow] (input) -- (lstm);
        \draw [arrow] (lstm) -- (fcn1);
        \draw [arrow] (fcn1) -- (fcn2);
        \draw [arrow] (fcn2) -- (output);
    \end{tikzpicture}
    \caption{The architecture of the neural network model from input to output. Each block represents a stage in the process, with arrows indicating the flow of data.}
    \label{fig:lstm_model}
\end{figure*}
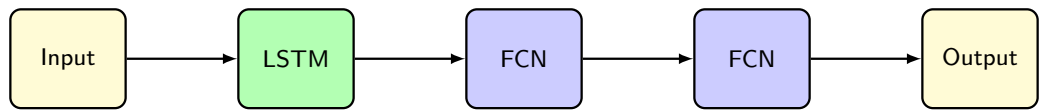

The development of the LSTM model adopts the manner of offline development and online deployment. The historical data stored by the platform is used for the model's training and testing. After the trained model meets the training criteria, it will be deployed in the platform as a container to work online.
Historical data includes two parts: the prepared and engineered data and corresponding actions. Actions here are operations recorded by the engineers manually when the QoS is lower than the threshold and result in disconnection.

The next anomaly must be forecasted according to the hidden pattern of the historical network and CPE state records, so a reasonable solution is to formalize a sequence prediction task, where multiple discrete records on sequential timestamps will be cascaded as one training sequence. Then these training sequences are labelled by using manual operations. The training sequence is labeled as `0’ if there is an action needed, otherwise, the label is `1’.

The features in Table~\ref{tab:engineered_features_full}, are some of the taken features in this study \footnote{\label{datadiscloserfootnote}\Ahmad{ This paper presents a complete overview of the AI-driven anomaly detection system, including data handling, platform architecture, model design, and deployment in a live offshore O-RAN environment. While the feature engineering process and input transformation methods are addressed, their coverage is limited due to commercially sensitive content, proprietary engineering practices and client-specific implementations. As a result, detailed result interpretation and feature-level analysis are discussed only at a high level to maintain transparency without disclosing proprietary  information.}}. Due to the feature engineering, we were able to replace some less important data like latitude, and longitude after using them to calculate more important features such as distance from all cells. The recording interval of fed data is 5 seconds and we used two months of records after being stored and then processed in our cloud-native data platform. The data pre-processing process is illustrated in Fig.~\ref{fig:raw_data pre-processing}.

First, for the given raw dataset, there are some records missing. So, the process needs to pad these missing values and remove the records that are obviously out of the normal range. Then as a step of de-noising, the data is downsampled to $1/N$ of the initial one. Then assemble every $L$ single feature vector as one training sequence. Lastly, the training sequence is labeled. It is worth mentioning that the criteria of labeling, are the manual actions taken by the engineer, which may lag over the first occurrence of the anomaly of $5$ to $20$ minutes. So, $10$ minutes is taken as the average delay applied to data labeling. The training sequence will be a matrix with size $1\times250\times L$, and the labels will be binary values. In this use case, after validation, $N=2$ and $L=6$ are reasonable options.

The architecture of the NN model is depicted in Fig.~\ref{fig:lstm_model}, wherein the LSTM module inherits from the LSTM function in Pytorch and the input size equals $1\times250\times 6$, hidden size equals 2. Two linear layers follow the LSTM module~\footnote{A tutorial of LSTM with detailed architecture and mathematical explanation can be found in~\cite{staudemeyer2019understanding}.}, and the output is the indicator of taking action or not. The last layer adopts the activation function 'sigmoid'.

\begin{figure}
    \centering
    \includegraphics[width=0.48\textwidth]{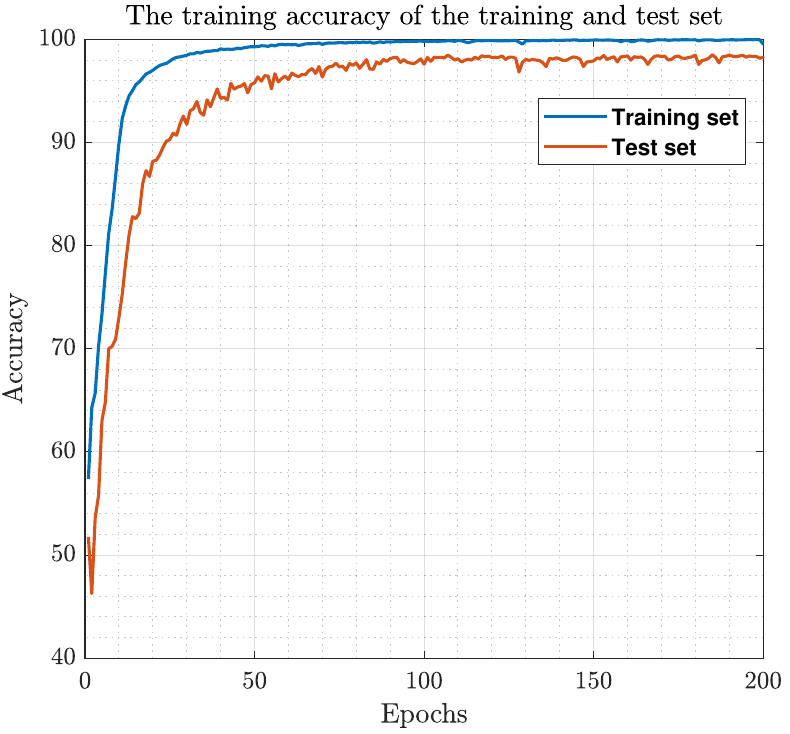}
    \caption{The model accuracy of the training and test set.}
    \label{fig:lstm_training_curve}
\end{figure}

For the model's training, the training/validation dataset is divided according to the 80\%/20\% rule. The loss function is the binary cross entropy (BEC)~\cite{PyTorchBCELoss}. 
The training batch size is $2048$ and Adam is taken as the optimizer. In the batch sampling process, the weighted batch sampler is used because this dataset is an imbalanced dataset, which means that labels '$1$'s are far more than labels '$0$'s. The above training dataset is used to train the LSTM model, and the test set is used to evaluate the trained model. In the training stage, the training data is shuffled before batch sampling. Fig.~\ref{fig:lstm_training_curve} shows the model's accuracy plots in the training and test. The convergence is reached after $100$ epochs.

\subsection{Model deployment and operational behaviour}

\Ahmad{The LSTM-based anomaly detection model was deployed as a containerized microservice within the Boldyn cloud-native data analytics platform. The model was developed and validated in the application layer, then dockerized and integrated into the DevOps pipeline. It is hosted in the processing layer, where it operates as part of the production environment. The microservice receives a continuous stream of real-time telemetry data from multiple sources, including CPE and O-RAN network functions, via the platform’s data streaming layer. The model processes incoming time-series data to identify patterns indicative of modem performance degradation.}
\begin{figure}
    \centering
    \includegraphics[width=1\linewidth]{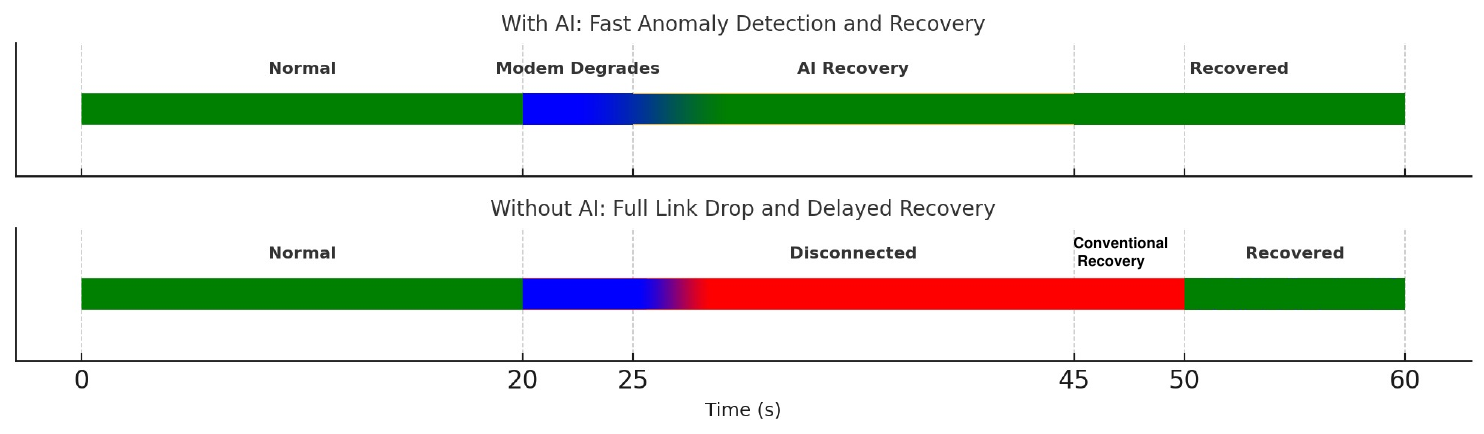}
    \caption{Anomaly detection operational behaviour impact compared to the conventional healing systems}
    \label{fig:operational_behaviour}
\end{figure}

\Ahmad{Once an anomaly is detected, the system immediately triggers a predefined control action, such as restarting or reconfiguring the modem, using available network APIs. The entire process, from detection to execution, is automated and designed for near real-time responsiveness.} \Ahmad{Fig.~\ref{fig:operational_behaviour} illustrates the expected operational behavior of the system under two conditions. In the top timeline (with AI, using LSTM), modem degradation is detected early (blue segment), and an automatic recovery action is taken (green transition) before a critical failure occurs. In contrast, the lower timeline (without AI, using conventional methods) where modem degradation progresses unnoticed into a full disconnection (red segment), followed by delayed manual recovery after alarm notifications. This diagram illustrate how the deployed system is architected to intervene earlier and avoid service impact through continuous anomaly monitoring and response automation.}

\subsection{Result discussion and operational effectiveness}

\Ahmad{The LSTM-based anomaly detection model was validated using live telemetry from an offshore O-RAN deployment involving dual-modem CPE devices operating under dynamic maritime conditions. Table ~\ref{table: real world results} summarizes weekly average network KPIs before and after the model's integration into the Boldyn cloud-native platform. Following deployment, the system demonstrated substantial gains in service resilience: single-modem disconnections fell from 170 to 20 per week, while dual-modem dropouts decreased from 25 to just 1. Manual network interventions were virtually eliminated, replaced by automated actions. Average response latency to degradations and disconnections shrank from 30 minutes to approximately 10 seconds.}
\Ahmad{In many of these cases, the AI model intervened before full service was lost, helping prevent modem failures that would have previously required human action or resulted in complete UE disconnection.}
\Ahmad{Prior to adopting the LSTM model, anomaly detection was handled by a rule-based conventional system crafted by radio engineers and SMEs from the experience and the observed cases faced before the disconnection. These rules used fixed limits and set conditions that worked for known problems but couldn’t adapt to new or unexpected issues. In real-world use, the system often failed to detect complex or unexpected failures, since creating rules for every possible case is extremely difficult. In contrast, the LSTM model was trained on past data and can recognize different types of failures by learning patterns over time. It was able to spot small signal issues and short-term changes that weren’t programmed in advance, leading to faster and more accurate responses.}

\begin{figure}
    \centering
    \includegraphics[width=0.9\linewidth]{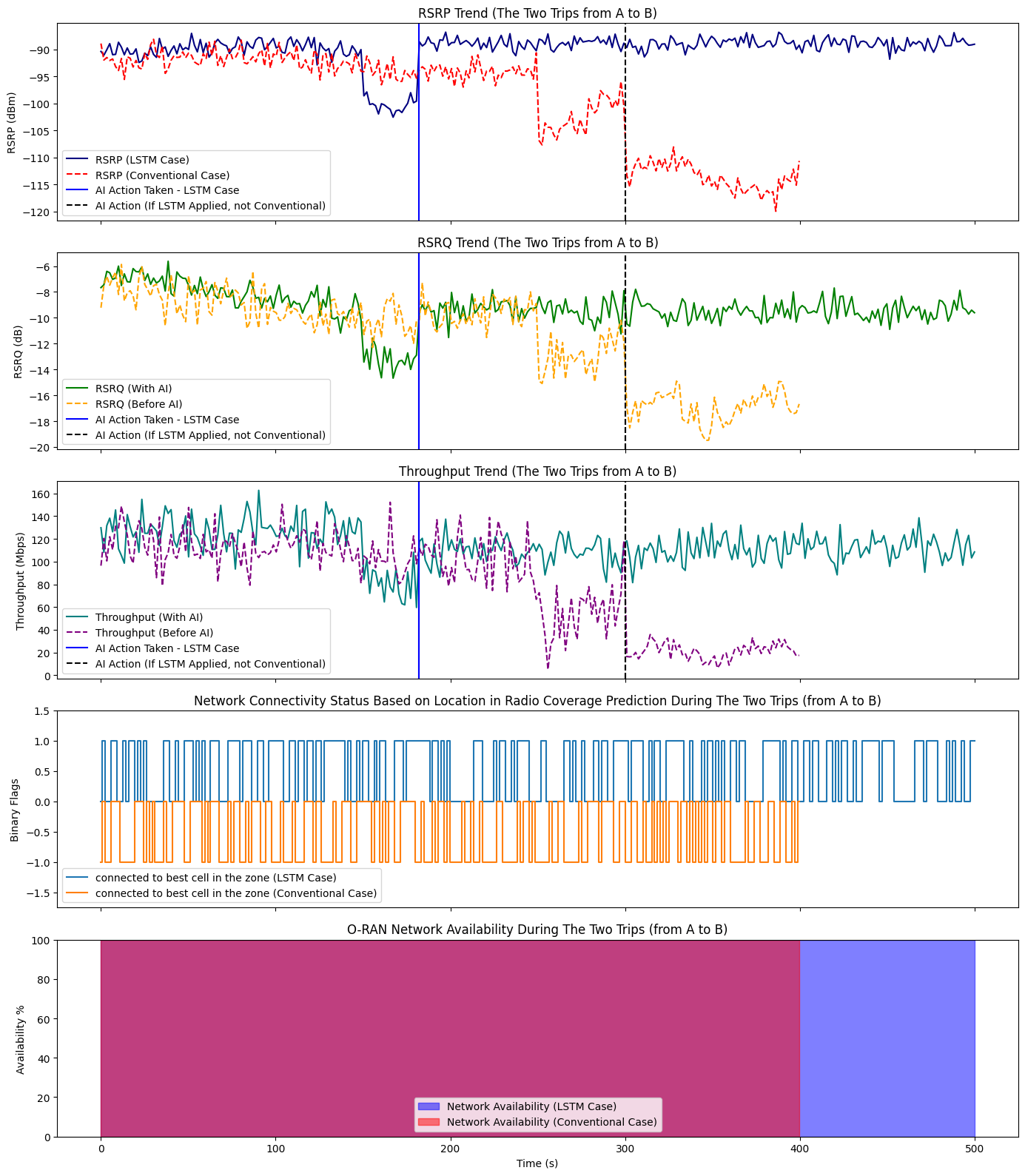}
    \caption{\textbf{Network Performance Before vs. After AI Deployment (Trip A to B).}
The chart compares two trips on the same route, one from history before AI applied (dotted lines) using rule-based logic, and one after deploying an LSTM model (solid lines). With AI, the system detected issues early and prevented disconnection, while the legacy setup failed to react in time, leading to a full drop.}
    \label{fig:trip_A_2_B}
\end{figure}
\Ahmad{To analyse the system’s results more closely, a route-specific comparison provides additional context. Fig.~\ref{fig:trip_A_2_B} shows telemetry data from the same vessel journey (Trip A to B) recorded at two different times. The first is from historical data, before the LSTM model was deployed, when only the conventional rule-based method was used. The second is from after the LSTM model was deployed. The data covers an 8-minute window and is plotted using dotted lines for the conventional system and solid lines for the LSTM-based system. In the LSTM-enabled case, the model detects the anomaly early (marked by the blue vertical line) and responds quickly, stabilising both RSRP, RSRQ and throughput. In contrast, the conventional system fails to detect the issue, which leads to a full disconnect. If the data from the conventional case had been processed by the LSTM model, it would have triggered an action earlier (marked by the black vertical dotted line). This highlights how the LSTM model performs better at detecting issues compared to the traditional system.}

\Ahmad{The last two charts in Fig.~\ref{fig:trip_A_2_B} show that looking only at network-side data or only at CPE data is not enough to understand the full picture. In this case, the network was working well, and there were no issues inside the vessel's local system. Also, the connectivity chart shows that the device was not always connected to the best cell most of the time. This is expected during vessel movement, especially between coverage sectors and outside the line-of-sight of the cell. In maritime environments, signals can reflect off the water or other surfaces, and sometimes the best signal comes from a tower that is not directly in front of the vessel.}

\Ahmad{These results (see footnote\textsuperscript{\ref{datadiscloserfootnote}}) show how complex the O-RAN environment can be. The conditions change often, coverage shifts from place to place, and many factors outside the system can affect performance. The LSTM model works well in this setting because it takes in many types of information, such as connection flags, mobility changes, signal levels, distance, throughput, and service availability and many more. By learning from this combined data, it can notice small changes in behavior that other systems would miss. This helps the system make better decisions, especially in large and mixed deployments where one data source alone cannot explain what is happening.}

\begin{table*}
\centering
\caption{Network performance before and after model deployment (measured weekly)}
\label{table: real world results}
\scalebox{0.88}{
\begin{tabular}{lccccccc}
\toprule
 & \textbf{One Modem} & \textbf{Two Modems} & \textbf{Network} & \textbf{Manual} & \textbf{Proactive} & \textbf{Response}  \\
 & \textbf{Disconnected} & \textbf{Disconnected} & \textbf{Disconnectivity} & \textbf{Actions Taken} & \textbf{Actions Taken} & \textbf{Time}  \\ \midrule
 \textbf{Before LSTM} & 170 & 25 & 25 & 25 & 0 & 30 mins  \\
 \textbf{After LSTM} & 20 & 1 & 1 & 1 & 170 & 10 sec  \\ \bottomrule
\end{tabular}}
\end{table*}

It is evident that connectivity significantly improved due to proactive measures taken when the model identifies potential connection drops. This has effectively prevented simultaneous loss of connection for both modems. Occasionally, such drops occur concurrently due to weather conditions or specific spatial factors, albeit infrequently and under unique circumstances.

Due to the innovative nature of this research and the emerging O-RAN technology, there is currently no publicly available dataset that closely matches the specific requirements of our study. Consequently, the data used in this research is proprietary and was collected from real-world deployments by Boldyn Networks. This limitation underscores the need for future efforts to develop and share standardized datasets to facilitate broader validation and comparison of AI models in similar contexts. 

All of this was possible because of the cloud-native data platform we built and the AI development Pyramid   standards we proposed. It allowed us to bring all the data together, support the development of the AI model, and run the system smoothly. This platform also makes it easier to add more complex systems in the future when needed.

\section{Discussions and future works}
\label{sec:discussions}

\begin{figure*}   
    \subfloat[\label{fig:Data1}]{
      \begin{minipage}[t]{0.5\linewidth}
        \centering 
        \includegraphics[width=3.2in]{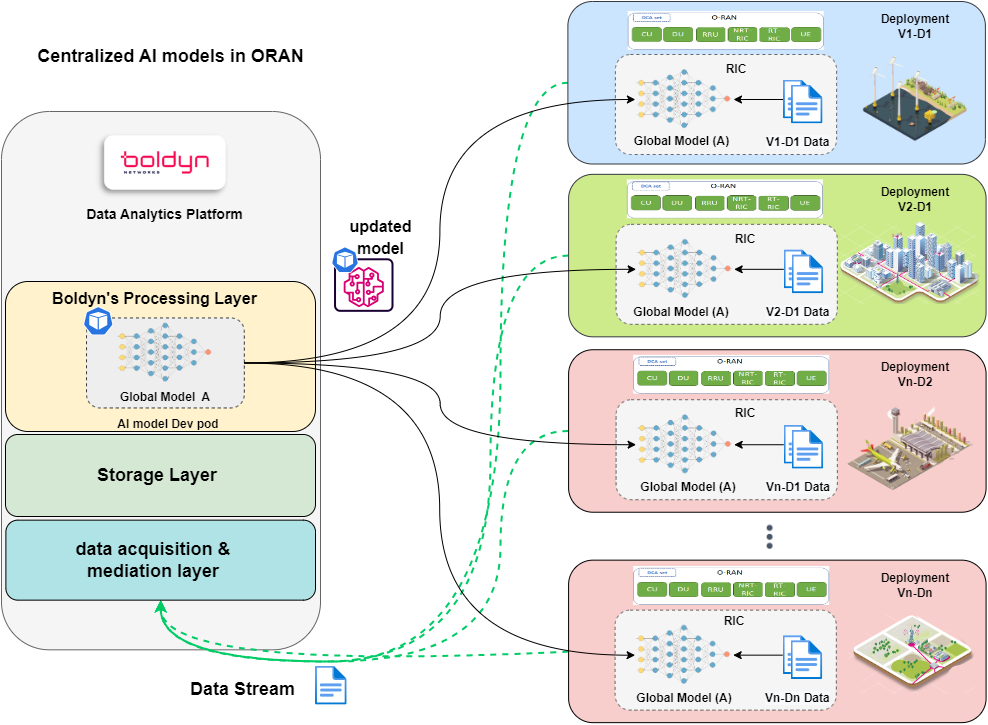}   
      \end{minipage}%
      }
        \subfloat[\label{fig:Data2}]{
      \begin{minipage}[t]{0.5\linewidth}   
        \centering   
        \includegraphics[width=3.2in]{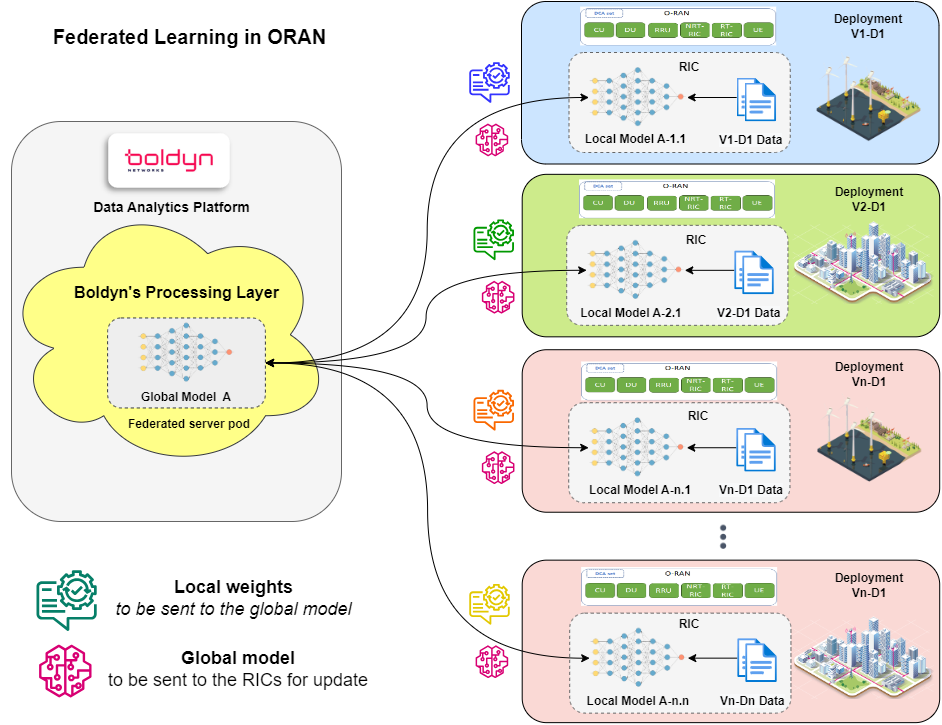}   
      \end{minipage} 
      }
      \caption{(a) This figure shows the illustration of the centralized AI model where the data from all the vendor systems and the network are centralized in the platform and then the AI model is trained, tested and validated in each specific O-RAN vendor. (b) present the potential federated learning aproach where the data is kept in each vendor system and the AI model is trained there and the global parameters are only shared with other deployments. this case is important when privacy is key.
      } \label{fig:Interpolation for data 1 and 2}
      \vspace{-0.2cm}
\end{figure*} 

\Ahmad{The deployment of the LSTM-based anomaly detection model within the Boldyn cloud-native platform has demonstrated clear benefits in enhancing connectivity, reducing service interruptions, and minimizing manual interventions. However, the current implementation is centered around a centralized training model, where raw telemetry from multiple offshore deployments is streamed to a shared processing layer for training and validation. While this model has proven effective in controlled, homogeneous environments, its scalability becomes more challenging in distributed, multi-tenant networks with differing operational policies, vendors, and data handling requirements.}
\Ahmad{A key consideration for future work is the need to support diverse client ecosystems, each potentially subject to unique privacy constraints, regulatory obligations, or internal governance rules. Transferring raw data from these domains into a centralized training pipeline may not always be permissible or optimal. Additionally, deployment-specific conditions (e.g., vendor configurations, physical topologies, local anomalies) may reduce the transferability of centrally trained models, necessitating more context-aware learning mechanisms.}

\Ahmad{To address these challenges, we propose extending the platform to support the decentralized learning framework, such as federated learning (FL). As illustrated in Fig.~\ref{fig:Data1}, this architecture would allow each deployment, regardless of vendor or location, to train its own local model on-site. Only model parameters or learned representations would be aggregated at the processing layer, avoiding the need to expose sensitive data. This approach is well-aligned with both data sovereignty requirements and the practical need for localized model specialization. By contrast, the current architecture (Fig.~\ref{fig:Data2}) requires central collection and processing of all telemetry, which becomes less sustainable as the network footprint grows.}

\Ahmad{The federated model also offers operational efficiencies: by enabling deployments to adapt to their own evolving conditions, it reduces the frequency of centralized retraining and better accommodates real-world heterogeneity. Boldyn’s platform already includes the modular components necessary to support such a transition, namely distributed inference pipelines, secure communication layers, and model versioning capabilities.}

\Ahmad{In future work, we plan to: (1) Prototype and benchmark FL-based training across representative offshore deployments; (2) Evaluate model performance consistency across heterogeneous vendor configurations; (3) Investigate synchronization strategies and update aggregation policies under bandwidth-constrained conditions; (4) Define compliance checkpoints for deployments governed by strict data residency and audit requirements.}

\Ahmad{These steps aim to evolve the platform from a performant centralized solution to a more resilient, privacy-aligned, and scalable AI framework, which is better suited to the multi-operator, multi-region nature of emerging O-RAN ecosystems.}

\section{Conclusions}
\label{sec:conclusion}
The radio access network (RAN) plays a critical role in modern telecom infrastructure, currently evolving towards disaggregated and open architectures such as O-RAN. These advances and innovations unlock opportunities for intelligent, data-driven applications to enhance network reliability and operational autonomy. However, the operation of O-RAN networks poses numerous research and engineering challenges due to immature real-world practices and complexities in managing data and applications across diverse vendor systems.
To address these challenges, Boldyn Networks developed a novel AI-driven, cloud-native data analytics platform. Tested with advanced LSTM models for real-time anomaly detection, the platform significantly improves operational efficiency and enhances customer experience. By leveraging DevOps practices and tailored data lakehouse architectures for AI applications, the platform exemplifies sophisticated data engineering strategies.
The deployment of this platform in an offshore O-RAN network demonstrated significant improvements in connectivity and operational efficiency, validating the model’s effectiveness \PL{with more than 90\% of connectivity issues resolved in run-time by the LSTM model.} 
However, the reliance on proprietary data highlights the need for standardized datasets to facilitate broader validation and comparison of AI models. \PL{Future research should explore the scalability of such AI-driven solutions across diverse, multi-vendor network environments. Implementing a decentralized platform, such as federated learning (FL) platform, could ensure consistent AI model performance while preserving data privacy across different regions and system configurations.}
This platform demonstrates significant potential for advancing in-RAN AI development. We aim to contribute to the community’s understanding and implementation of complex challenges in this domain, fostering innovations and improvements.

\section*{CRediT authorship contribution statement}

\begin{itemize}
    \item \textbf{Abdelrahim Ahmad:} Conceptualization, Data curation, Formal analysis, Investigation, Methodology, Software, Validation, Visualization, Writing - original draft.
    \item \textbf{Peizheng Li:} Conceptualization, Data curation, Formal analysis, Investigation, Methodology, Software, Validation, Visualization, Writing - original draft.
    \item \textbf{Robert Piechocki:} Project administration, Supervision, Writing - review \& editing.
    \item \textbf{Rui Inacio:} Conceptualization, Methodology, Validation, Writing - review \& editing.
\end{itemize}

\section*{Funding}
This work is a continuation of the research conducted within the Innovate UK/CELTIC-NEXT European collaborative project AIMM (AI-enabled Massive MIMO), which partially funded this research.

\section*{Declaration of competing interest}
The authors declare that they have no known competing financial interests or personal relationships that could have appeared to influence the work reported in this paper.

\section*{Acknowledgment}

The authors would like to sincerely thank the following individuals from Boldyn Networks for their invaluable contributions to this paper: Sean Keating, Chief Technology Officer UK \& Ireland, for his managerial support; Andrew Conway, Group Director Technology Strategy, Donal O’Sullivan, Head of Product Innovation, and David Kinsella, RAN Solutions Architect, for their technical review of the paper; and Menglin Yao, Data \& Software Engineer, and Michael Waldron, DevOps Engineer, for their platform operation and technical support. Their review, constructive comments, and support were instrumental in the development and completion of this work.

\bibliographystyle{ieeetr}
\bibliography{cas-refs}

\begin{thebibliography}{10}

\bibitem{OpenRANyang}
M.~Yang {\em et~al.}, ``{OpenRAN: A Software-Defined Ran Architecture via Virtualization},'' {\em SIGCOMM Comput. Commun. Rev.}, vol.~43, p.~549–550, aug 2013.

\bibitem{li2023digital}
P.~Li {\em et~al.}, ``{A Digital Twin of the 5G Radio Access Network for Anomaly Detection Functionality},'' in {\em Proc. of IEEE ICNP}, IEEE, 2023.

\bibitem{frcmn.2023.1127039}
M.~Alavirad {\em et~al.}, ``{O-RAN architecture, interfaces, and standardization: Study and application to user intelligent admission control},'' {\em Front. Commun. Netw.}, vol.~4, 2023.

\bibitem{polese2023understanding}
M.~Polese {\em et~al.}, ``{Understanding O-RAN: Architecture, Interfaces, Algorithms, Security, and Research Challenges},'' {\em IEEE Commun. Surv. Tutor.}, 2023.

\bibitem{10329948}
A.~M. Nagib, H.~Abou-Zeid, and H.~S. Hassanein, ``{Safe and Accelerated Deep Reinforcement Learning-Based O-RAN Slicing: A Hybrid Transfer Learning Approach},'' {\em IEEE J. Sel. Areas Commun.}, vol.~42, no.~2, pp.~310--325, 2024.

\bibitem{singh2020evolution}
S.~K. Singh, R.~Singh, and B.~Kumbhani, ``{The Evolution of Radio Access Network Towards Open-RAN: Challenges and Opportunities},'' in {\em Proc. of IEEE WCNCW}, pp.~1--6, IEEE, 2020.

\bibitem{armbrust2021lakehouse}
M.~Armbrust {\em et~al.}, ``{Lakehouse: A New Generation of Open Platforms that Unify Data Warehousing and Advanced Analytics},'' in {\em Proc. of CIDR}, vol.~8, p.~28, 2021.

\bibitem{ebert2016devops}
C.~Ebert {\em et~al.}, ``{DevOps},'' {\em IEEE Software}, vol.~33, no.~3, pp.~94--100, 2016.

\bibitem{faisal_2021}
Faisal, ``{RAN Vs Cloud RAN Vs VRAN Vs O-RAN: A Simple Guide!},'' Apr 2021.

\bibitem{O_RAN2023}
{O-RAN Alliance}, ``{O-RAN.WG4.CTI-TCP.0-R003-v04.00: Cooperative Transport Interface - Transport Control Plane Specification},'' tech. rep., O-RAN Alliance, 2023.
\newblock Accessed: 25 July 2024.

\bibitem{O_RAN2024}
{O-RAN Alliance}, ``{O-RAN.WG1.OAD-R003-v12.00: O-RAN Architecture Description},'' tech. rep., O-RAN Alliance e.V., 2024.
\newblock Accessed: 25 July 2024.

\bibitem{10453148}
A.~Aijaz {\em et~al.}, ``{Open RAN for 5G Supply Chain Diversification: The BEACON-5G Approach and Key Achievements},'' in {\em Proc. of IEEE CSCN}, pp.~1--7, 2023.

\bibitem{BONATI2023109502}
L.~Bonati, M.~Polese, S.~D’Oro, S.~Basagni, and T.~Melodia, ``Openran gym: Ai/ml development, data collection, and testing for o-ran on pawr platforms,'' {\em Computer Networks}, vol.~220, p.~109502, 2023.

\bibitem{Polese10024837}
M.~Polese, L.~Bonati, S.~D’Oro, S.~Basagni, and T.~Melodia, ``Understanding o-ran: Architecture, interfaces, algorithms, security, and research challenges,'' {\em IEEE Communications Surveys \& Tutorials}, vol.~25, no.~2, pp.~1376--1411, 2023.

\bibitem{Yeh10335921}
S.-P. Yeh, S.~Bhattacharya, R.~Sharma, and H.~Moustafa, ``Deep learning for intelligent and automated network slicing in 5g open ran (oran) deployment,'' {\em IEEE Open Journal of the Communications Society}, vol.~5, pp.~64--70, 2024.

\bibitem{Nagib10329948}
A.~M. Nagib, H.~Abou-Zeid, and H.~S. Hassanein, ``Safe and accelerated deep reinforcement learning-based o-ran slicing: A hybrid transfer learning approach,'' {\em IEEE Journal on Selected Areas in Communications}, vol.~42, no.~2, pp.~310--325, 2024.

\bibitem{Hakan10012789}
H.~Erdol, X.~Wang, P.~Li, J.~D. Thomas, R.~Piechocki, G.~Oikonomou, R.~Inacio, A.~Ahmad, K.~Briggs, and S.~Kapoor, ``Federated meta-learning for traffic steering in o-ran,'' in {\em 2022 IEEE 96th Vehicular Technology Conference (VTC2022-Fall)}, pp.~1--7, 2022.

\bibitem{11005418}
H.~Li, P.~Li, K.~D. Assis, J.~M. Parra, A.~Aijaz, S.~Yan, and D.~Simeonidou, ``Netmind+: Adaptive baseband function placement with gcn encoding and incremental maze-solving drl for dynamic and heterogeneous rans,'' {\em IEEE Transactions on Network and Service Management}, pp.~1--1, 2025.

\bibitem{BASARAN2025111145}
O.~T. Basaran and F.~Dressler, ``Xainomaly: Explainable and interpretable deep contractive autoencoder for o-ran traffic anomaly detection,'' {\em Computer Networks}, vol.~261, p.~111145, 2025.

\bibitem{Xavier10279349}
B.~M. Xavier, M.~Dzaferagic, D.~Collins, G.~Comarela, M.~Martinello, and M.~Ruffini, ``Machine learning-based early attack detection using open ran intelligent controller,'' in {\em ICC 2023 - IEEE International Conference on Communications}, pp.~1856--1861, 2023.

\bibitem{recentadvancement23218792}
M.~Q. Hamdan {\em et~al.}, ``{Recent Advances in Machine Learning for Network Automation in the O-RAN},'' {\em Sensors}, vol.~23, no.~21, 2023.

\bibitem{erdol2022federated}
H.~Erdol {\em et~al.}, ``{Federated Meta-Learning for Traffic Steering in O-RAN},'' in {\em Proc. of IEEE VTC2022-Fall}, pp.~1--7, IEEE, 2022.

\bibitem{li2024netmind}
H.~Li, P.~Li, K.~D. Assis, A.~Aijaz, S.~Shen, R.~Nejabati, S.~Yan, and D.~Simeonidou, ``Netmind: Adaptive ran baseband function placement by gcn encoding and maze-solving drl,'' in {\em 2024 IEEE Wireless Communications and Networking Conference (WCNC)}, pp.~1--6, IEEE, 2024.

\bibitem{ANS10335921}
S.-P. Yeh {\em et~al.}, ``{Deep Learning for Intelligent and Automated Network Slicing in 5G Open RAN (ORAN) Deployment},'' {\em IEEE Open J. Commun. Soc.}, vol.~5, pp.~64--70, 2024.

\bibitem{kundu2024towards}
L.~Kundu, X.~Lin, and R.~Gadiyar, ``{Towards Energy Efficient RAN: From Industry Standards to Trending Practice},'' {\em arXiv preprint arXiv:2402.11993}, 2024.

\bibitem{ric}
B.~Balasubramanian {\em et~al.}, ``{RIC: A RAN Intelligent Controller Platform for AI-Enabled Cellular Networks},'' {\em IEEE Internet Computing}, vol.~25, no.~2, pp.~7--17, 2021.

\bibitem{AI-enabled_O-RAN}
B.-S.~P. LinI, ``{Toward an AI-enabled O-RAN-based and SDN/NFV-driven 5G\& IoT network era},'' {\em Netw. Commun. Technol.}, vol.~6, no.~1, p.~6, 2021.

\bibitem{Soltani9881863}
S.~Soltani {\em et~al.}, ``{Can Open and AI-Enabled 6G RAN Be Secured?},'' {\em IEEE Consum. Electron. Mag.}, vol.~11, no.~6, pp.~11--12, 2022.

\bibitem{li2022transmit}
P.~Li {\em et~al.}, ``{Transmit Power Control for Indoor Small Cells: A Method Based on Federated Reinforcement Learning},'' in {\em Proc. of IEEE VTC2022-Fall}, pp.~1--7, IEEE, 2022.

\bibitem{ITU2024}
{International Telecommunication Union}, ``{ITU-T Recommendation M.3010: Principles for a Telecommunications Management Network},'' tech. rep., International Telecommunication Union (ITU), 2024.
\newblock Accessed: 25 July 2024.

\bibitem{Peizheng9931127}
P.~Li {\em et~al.}, ``{RLOps: Development Life-Cycle of Reinforcement Learning Aided Open RAN},'' {\em IEEE Access}, vol.~10, pp.~113808--113826, 2022.

\bibitem{GitLabGitOps2024}
{GitLab}, ``{GitOps: A Comprehensive Guide},'' 2024.
\newblock Accessed: 25 July 2024.

\bibitem{AHMED201619}
M.~Ahmed, A.~{Naser Mahmood}, and J.~Hu, ``{A Survey of Network Anomaly Detection Techniques},'' {\em Journal of Network and Computer Applications}, vol.~60, pp.~19--31, 2016.

\bibitem{staudemeyer2019understanding}
R.~C. Staudemeyer and E.~R. Morris, ``Understanding lstm--a tutorial into long short-term memory recurrent neural networks,'' {\em arXiv preprint arXiv:1909.09586}, 2019.

\bibitem{PyTorchBCELoss}
{PyTorch Contributors}, {\em BCELoss}.
\newblock PyTorch, 2023.

\end{thebibliography}

\end{document}